\documentclass[letterpaper, submit]{AAS}			% to submit to JAS
\usepackage{bm}
\usepackage{amsmath}
\usepackage{amssymb}
\usepackage[colorlinks=true, pdfstartview=FitV, linkcolor=black, citecolor= black, urlcolor= black]{hyperref}
\usepackage{comment}
\usepackage{empheq}
\usepackage{overcite}
\usepackage{float}
\usepackage{footnpag}			      	% make footnote symbols restart on each page

\usepackage{algorithm}
\usepackage{algpseudocode}
\usepackage{fancyhdr}
\usepackage{graphicx}
\usepackage{subcaption}

\PaperNumber{25-428}

\begin{document}

\title{Probabilistic Methods for Initial Orbit Determination and Orbit Determination in Cislunar Space\thanks{DISTRIBUTION A: Approved for public release; distribution is unlimited. Public affairs approval \#AFRL-2026-0779.}}

\author{Ishan P. Paranjape\thanks{Graduate Assistant - Research, Department of Aerospace Engineering, Texas A\&M University, 710 Ross St., College Station, TX 77843.}, Tarun Hejmadi\thanks{Graduate Assistant - Research, Department of Aerospace Engineering, Texas A\&M University, 710 Ross St., College Station, TX 77843.}, and 
Suman Chakravorty\thanks{Professor, Department of Aerospace Engineering, Texas A\&M University, 710 Ross St., College Station, TX 77843.},
}

\maketitle{} 	

% \thanks{DISTRIBUTION A: Approved for public release; distribution is unlimited. Public Affairs release approval \#_______.}

\begin{abstract}
In orbital mechanics, Gauss's method for orbit determination (OD) is a popular, minimal assumption solution for obtaining the initial state estimate of a passing resident space object (RSO). Since much of the cislunar domain relies on three-body dynamics, a key assumption of Gauss's method is rendered incompatible, creating a need for a new, minimal assumption method for initial orbit determination (IOD). In this work, we present a framework for short and long term probabilistic target tracking in cislunar space which produces an initial state estimate with as few assumptions as possible. Specifically, we propose an IOD method involving the kinematic fitting of several series of noisy, consecutive ground-based observations. Once a probabilistic initial state estimate in the form of a particle cloud is formed, we apply the powerful Particle Gaussian Mixture (PGM) Filter to reduce the uncertainty of our state estimate over time. This combined IOD/OD framework is demonstrated for several classes of trajectories in cislunar space and compared to better-known filtering frameworks.

\end{abstract}

\section{Introduction}\label{sec:1intro}
% \textcolor{red}{Please number all sections, this is highly confusing.} \textcolor{blue}{I found the Springer Nature template and will migrate all content to it after the prelims and adjust section references accordingly.}
One of the main objectives of the National Cislunar Science and Technology Strategy is the extension of space situational awareness (SSA) capabilities into cislunar space. Specifically, there exists a need to identify and detect colliding objects, potentially dangerous spacecraft operations, or space weather elements in order to ensure safe operation of spacecraft within this domain.\cite{ussf2020}  The volume of the cislunar domain -- encompassing the full $4\pi$ steradian volume in which three-body dynamics including the Earth and Moon are modeled -- is nearly a thousand-fold that of the volume of orbits below the geocentric (GEO) limit. Furthermore, several Resident Space Objects (RSOs) -- also referred to as targets more generally -- move within this domain, so effective cislunar target tracking becomes a crucial problem. The rise of lunar missions such as India's Chandrayaan and China's Chang'e-5 makes it even more imperative to facilitate space traffic coordination efforts and ensure the safe operation of spacecraft in this domain. One of the ways we can ensure this safe operation is by using initial orbit determination (IOD) and orbit determination (OD) to initialize and maintain custody of these RSOs.

The goal of IOD for tracking a single target is to combine several measurements of a single RSO and then to fit a state vector through these measurements.\cite{gooding1997} While Gauss's method of orbit determination remains one of the most well-known IOD methods for the two-body problem, more specialized IOD methods such as constrained admissible region (CAR) and probabilistic admissible region (PAR) have been suggested in the literature. \cite{taff1979, deMars2013, kelecy2013, hussein2014, hussein2014.2} One set of researchers utilizes PAR to generate initial estimates of a target within regimes up to geosynchronous orbit (GEO), focusing on two-body dynamics and Keplerian motion.\cite{mishra2024} More specifically, their IOD framework involves postulating statistics about four of the six major orbital elements (corresponding to some orbital plane) and generating multiple possibilities of the target state using those statistics.\cite{mishra2024} Along with two angular measurements, an initial state estimate is formulated. Another set of researchers utilizes PAR to generate a similar set of initial state estimates in the cislunar domain.\cite{bolden2022, griggs2023} Since traditional orbital elements cannot be used in the cislunar domain, where effects of the three-body problem dominate and only small sub-collections of orbits are elliptic in nature, the authors rely upon postulated statistics of the range utilizing experimental data about light intensity curves, sensor movement constraints for angular rates, and the singular constant of integration (Jacobi's constant) produced by the equations of motion of the circular-restricted three-body problem (CR3BP).\cite{hejduk2013} 

Although most probablistic IOD methods are expected to yield an accurate initial state estimate, such an estimate is not always precise. As a result, orbit determination (OD) becomes a critical component of target tracking. For probabilistic initial state estimates, OD implies the use of filters and uncertainty reduction. Generally, in cases of more compact, approximately Gaussian PDFs, an Extended Kalman Filter (EKF), Unscented Kalman Filter (UKF), or Ensemble Kalman Filter (EnKF) may be practical.\cite{kalmanBucy1961, crassidis2011, wan2000, julier2004, Evensen1994, evensen1995, evensen2003} For more highly nonlinear and chaotic dynamical systems, several techniques such as square root UKFs and adaptive Gaussian Mixture (AGM) filters, whose \textit{a priori} estimates modeled as Gaussian Mixture Models (GMMs), have been developed.\cite{merwe2003, wan2000, demars2013_entropy, liu2016, popov2023, grueso2025, Iannamorelli2025AdaptiveGM} In parallel, hybrid filters such as the UKF/particle filter have been created to combine the strengths of existing filters at either the propagation or update steps.\cite{raihanukfpf2018} The hybrid UKF/particle filter formulation in particular has led to the development of the Particle Gaussian Mixture (PGM) Filter.

The PGM Filter was developed to address two major issues in GMM based filters: 1) inflexibility of the number of components and weights during nonlinear propagation, and 2) the curse of dimensionality in the form of particle depletion \cite{raihan2016, raihan2018} This iterative filter consists of four key steps: particle propagation, clustering, measurement update, and resampling. The PGM Filter has been successfully demonstrated for the two-body problem, and is also applicable for multi-target tracking scenarios.\cite{mishra2024, mishra2023}. In parallel, the same set of researchers that extended the PAR technique to cislunar space utilize the PGM Filter for OD.\cite{bolden2022, griggs2023} Due to its robustness over multiple chaotic dynamical systems, our work shall retain the PGM Filter for the OD aspect of our target tracking framework.

While PAR has proven to be an effective method of generating a probabilistic initial state estimate in the cislunar domain, a major drawback of this method involves postulating several statistics and requires making numerous \textit{a priori} assumptions about the measurement tools, the ability to see an object, as well as target geometry. Furthermore, this work utilizes two different dynamic system models. Gauss's method for orbit determination remains a benchmark for IOD due to its limited set of assumptions -- the most important of which are Keplerian dynamics and sufficiently short cadences between measurements -- and the mathematical manipulations associated with obtaining an initial state estimate of an RSO up to the GEO regime by using three consecutive short-arc measurements. Gauss's method does not hold in the cislunar domain due to violation of the Keplerian (i.e. two-body) assumption. In this work, we showcase a combined, minimal-assumption framework for IOD and OD in the cislunar domain through kinematic fitting of our observations and using various filtering frameworks for target tracking, with a focus on the PGM Filter. For IOD, we utilize the fact that RSOs in the cislunar domain remain within the line of sight of an observer for extended periods of time, allowing us to observe the target's position over some initial period of time, fit a polynomial spline to the observations and determine the object's state after the initial period by differentiating to obtain the velocities and thereby the state. This process is repeated for different sets of observations obtained by perturbing the actual measurements with the sensor noise to obtain a particle cloud for the initial state which is then fitted using a Gaussian mixture model (GMM) giving us an initial state probability density function which is then tracked using the PGM filter.

The remaining sections of this article are as follows. First, we explain the nonlinear dynamics and measurement models used for initial orbit determination and subsequent target tracking. Then, we separately introduce these IOD and target tracking frameworks. Next, we present several different examples with our combined IOD/OD framework and compare the performance of our OD framework with other nonlinear filtering frameworks. We also demonstrate the utility of our PGM filter on an orbit in which other filtering frameworks will fail. Then, we discuss three ways to utilize range or velocity information to obtain an initial state estimate. We conclude with a discussion about some of the limitations of our IOD/OD framework.

\section{Cislunar IOD and Target Tracking Preliminaries}\label{sec:2prelims}

Effective target tracking in the cislunar domain encompasses two key features: an IOD method to generate an initial state estimate, and an OD method to continuously track and maintain custody of an RSO. In regimes such as LEO or GEO -- governable by Kepler's equations of motion -- ground-based observers can have only brief glimpses of RSOs due to the short target ranges relative to the target velocities. A single pass, defined as a continuous trajectory during which a single object in space is within a ground-based sensor's field of view, can last only a few minutes. The vastness of the cislunar domain results allows for targets further out to move much more slowly relative to the ground-based observer. A single pass within this domain lasts between 10-20 hours, limited only by the speed of Earth's rotation about its own axis. Due to the orders of magnitude difference between pass lengths between the cislunar and near-Earth domains, we are able to fit state vectors through several observations within a single pass to obtain an initial state estimate for some point in the middle of the pass. We dedicate this section to outlining the dynamics and measurement models which we utilize in this work that lend themselves to making the development of our IOD-OD framework possible. 

\subsection{Dynamics Model: Circular Restricted Three Body Problem}\label{subsec:cr3bp}

Beyond the GEO regime, modeling RSO dynamics with Keplerian motion or two-body dynamics becomes difficult due to the effects of massive gravitational bodies such as the Moon and the Sun. The addition of these large celestial bodies renders the use of Keplerian orbital elements obsolete. Nevertheless, multiple dynamics models exist for bodies in cislunar space, many of which are derived from the three-body problem.\cite{koon1999} The most popular model for explaining three-body dynamics, especially in cases in which the mass of one body (i.e. the spacecraft or RSO) is negligible compared the other two bodies (i.e. the Earth and the Moon) is the circular-restricted three-body problem (CR3BP).\cite{holzinger2021, schaub2003} 

CR3BP dynamics are expressed in a rotating reference frame centered around the joint center of mass of the two largest orbiting bodies (also referred to as the barycenter). These two large bodies (in our case, the Earth and the Moon) orbit about their joint center of mass in a circular motion. The RSO is generalized as a third body with negligible mass relative to the other two bodies, and one whose motion in this reference frame is relative to the positions of the Earth and the Moon at any given time. The equations of motion of the target with respect to the barycentric frame are given below.

\begin{subequations} \label{eq3:cr3bp}
    \begin{align}
        \ddot{x} = x + 2\dot{y} - \frac{(1-\mu)(x+\mu)}{r_1^3} - \frac{\mu x - \mu(1-\mu)}{r_2^3}\label{eq3a:cr3bp_X} \\
        \ddot{y} = y - 2 \dot{x} - \frac{(1-\mu)y}{r_1^3} - \frac{\mu y}{r_2^3}\label{eq3b:cr3bp_Y} \\
        \ddot{z} = \frac{(1-\mu)z}{r_1^3} - \frac{\mu z}{r_2^3} \label{eq3c:cr3bp_Z}
    \end{align}
\end{subequations}

In the equations above, $x$ and $y$ define the orbital plane of the Earth and the Moon, $r_1$ and $r_2$ define the distance between the target and the centers of masses of the Earth and Moon, respectively. $\mu$ is a mass parameter that defines the weights of the two major bodies relative to each other, and is assumed constant. The distance between the centers of the Earth and the Moon is non-dimensionalized to 1 n.d. for simplicity. As a result, the barycenter is located at distance of $\mu$ to the right of Earth on the $x-$axis and $1 - \mu$ to the left of the Moon along the same axis. The following table describes the conversion of non-dimensionalized quantities of mass, time, and distance expressed in commonly known units.

\begin{table}[h!]
    \centering
    \begin{tabular}{||c c||} 
     \hline
     Non-Dimensionalized Quantity & Conversion\\ [0.5ex] 
     \hline\hline
     Mass Ratio ($\mu$) & $1.2150582 \times 10^{-2}$\\ 
     \hline
     Time & $4.342$ days\\ 
     \hline
     Distance & $384400$ km. \\
     \hline
     Velocity & $1.0247$ km./s \\[1ex] 
     \hline
    \end{tabular}
    \caption{Non-dimensionalized quantities relevant to CR3BP dynamics, per \textit{Schaub \& Junkins 2003}\cite{schaub2003}}
    \label{t1:ndquantities}
\end{table}

The CR3BP dynamical system consists of five equilibrium solutions, known commonly as Lagrange points. The first three of these Lagrange points, labeled $L_1$, $L_2$, and $L_3$, lie along the Earth-Moon $x-$axis and are regarded as the points of unstable equilibrium. On the other hand, the $L_4$ and $L_5$ Lagrange points -- which form an equilateral triangle with the centers of the Earth and the Moon -- are regarded as the points of stable equilibrium.\cite{doedel2007, szebehely1969} However, probabilistic estimates centered near or around the $L_1$ to $L_3$ Lagrange points will undergo some of the most chaotic bifurcations and deformations over the shortest period of time.\cite{zimovan2017, zimovanspreen2020} We explore such trajectories later in this article.

% An RSO placed that those two points at zero velocity will stay at that point indefinitely. This type of orbit or initial condition may be used as a benchmark for testing the IOD/OD framework described in the proceeding section. (Reserved for dissertation)

\subsection{Measurement Model}\label{subsec:measModel}

Four important points exist within our CR3BP dynamics model: the barycenter \textit{B}, the center of the Earth \textit{E}, the center of the Moon \textit{M}, and an Earth-based observer or ground station \textit{O}. Let $\mathbf{r}_{BT}$ and $\mathbf{r}_{BO}$ represent position vectors defining the positions of the target and ground-based observer with respect to the barycenter in the common CR3BP reference frame. Then, the position of the target with respect to the observer in this barycentric reference frame $\mathcal{B}$ becomes
\begin{equation}\label{eq4:vectorAlgebra}
    \mathbf{r}_{OT}^{\mathcal{B}} = \mathbf{r}_{BT} - \mathbf{r}_{BO}.
\end{equation}

The sensor provides measurements in a topocentric reference frame specific to a point on Earth's surface. To account for the difference in reference frames between our measurement model and the dynamics model, we note that Eq. \ref{eq4:vectorAlgebra} may be decomposed into
\begin{equation}\label{eq5:boDecomp}
    \mathbf{r}_{BO} = \mathbf{r}_{BE} + \mathbf{r}_{EO}.
\end{equation}
A constant quantity in the barycentric reference frame, $\mathbf{r}_{BE}$ may be converted into the Earth-centered inertial (ECI) reference frame using the simple $R_3$ transformation given as
\begin{equation}\label{eq6:rotation}
    R_{CR3BP \rightarrow ECI} =
    \begin{bmatrix}
        cos(\omega t) & sin(\omega t) & 0 \\
        -sin(\omega t) & cos(\omega t) & 0 \\
        0 & 0 & 1
    \end{bmatrix}.
\end{equation}
In the CR3BP model, $\omega$ (the rotation rate) is non-dimensionalized to one and the transformation matrix becomes a function of the non-dimensionalized time as a result. By knowing the location of the observer and the specific instance of time, we can compute $\mathbf{r}_{EO}$ in the ECI reference frame using software tools. With $\mathbf{r}_{BE}$, $\mathbf{r}_{EO}$, and $\mathbf{r}_{EO}$ all in the ECI frame, it is now possible to construct an estimate for $\mathbf{r}_{OT}$ in the topocentric frame using another simple transformation. For our measurement model, we use a sensor-fixed, local NEU (North-East-Up) coordinate frame. We denote the components of $\mathbf{r}_{OT}$ in this NEU topocentric reference frame $\mathcal{T}$ as
\begin{equation}\label{eq7:posTopo}
    \mathbf{r}_{OT}^{\mathcal{T}} = [r_x, r_y, r_z].
\end{equation}

By pointing in the direction of the RSO, SSA telescopes and radars localize the azimuth (AZ) and elevation (EL) of the target. With the set of position coordinates of the RSO/target in the topocentric reference frame $\mathcal{T}$ by Eq. \ref{eq7:posTopo}, the AZ and EL angles may be computed as follows:
\begin{subequations} \label{eq8:AZ-EL}
    \begin{align}
        AZ = tan^{-1} (\frac{r_y}{r_x}) \label{eq8a:azimuth} \\
        EL = \frac{\pi}{2} - cos^{-1} (\frac{r_z}{\sqrt{{r_x}^2 + {r_y}^2 + {r_z}^2}}) \label{eq8b:elevation}
    \end{align}
\end{subequations}

To convert AZ and EL back into $[r_x, r_y, r_z]$ quantities, range information $\rho$ must be utilized, in which $\rho = \sqrt{r_x^2 + r_y^2 + r_z^2}$. Range may be inferred or indirectly measured using passive radio frequency (RF) measurement types such as Time Delay of Arrival (TDOA), One-Way Doppler (OWD), or a combination of the two, with one or more ground-based radar stations.\cite{griggs2023} Range information in cislunar space does not tend to be as reliable due to the inverse relationship between received signal power and range. We model this unreliability by assuming high, unbiased range measurement noise. The standard deviation for this noise is chosen to be some non-insignificant percentage of the true range. However, due to the high fidelity of electro-optical (EO) sensors, we assume that one standard deviation of angular errors for both AZ and EL is 1.5 arcsec. 

As we shall explain in the following section, range information will only be necessary for an initial state estimate. Beyond that, angles-only (i.e. AZ and EL) measurements will be sufficient for orbit determination. Furthermore, by sampling a large number of observations and observation sets from the noise statistics, many particles are expected to lie outside the cislunar domain -- defined by roughly twice the GEO distance from the center of the Earth up to several hundreds of thousands of kilometers. We assume that our target will always lie within the bounds of cislunar space at least during its first pass, and any particle found outside of this domain shall be filtered out for both IOD and OD. 

\section{Target Tracking Framework}\label{sec:3ttf}

Our minimal-assumption target tracking framework consists of an IOD framework and an OD framework that work sequentially. Our IOD framework creates an initial state estimate by utilizing range information and angular measurements. The OD framework, consisting of a PGM filter, subsequently helps reduce the uncertainty of our initial state estimate over a long period of time. In this section, we outline both the IOD and OD frameworks in detail.

\subsection{Initial Orbit Determination}\label{subsec:IOD}

Our minimal-assumption IOD framework derives inspiration from Gauss's method for orbit determination. Gauss's method derives an initial state estimate using three angles-only observations of an object of interest, orbiting in Keplerian motion.\cite{bmw1971, vallado2013} By making note of the observer position with respect to the center of the Earth and the slant range, Gauss's method utilizes Lagrange coefficients and the fact that times between consecutive observations are sufficiently small to obtain an initial state estimate (i.e. position and velocity) at the second timestep.\cite{curtis2014} Unfortunately, Gauss's method is not valid in cislunar space because the assumption of Keplerian motion no longer becomes valid. Due to the effect of another large body in the cislunar domain (i.e. the Moon), motion is no longer planar relative to an ECI reference frame, and the use of orbital elements for geometric PAR is no longer possible.\cite{mishra2024} In order to generate an initial state estimate for a target in cislunar space subject to three-body motion, a method independent of assumptions about the system dynamics must be utilized. To that end, we propose a method for deriving a full-state estimate from a full-position estimate, making use of angles-only measurements and minimal information about the slant range.

Due to the vast distances of cislunar space, it may be noted that a target of interest, once spotted by an observer, will remain in the observer's field of view (FOV) for a significantly long time. Accounting for the Earth's rotation, an object will reside in an observer's FOV for roughly 10-20 hours. A ground-based observer is then able to take tens of measurements throughout this 10-20 hour period pass, compared to the minutes-long passes of targets below the GEO limit. We can task an optical sensor to take angular measurements of a target and fuse these measurements with some range information at a sufficiently high frequency that we obtain a time-series data of the target position in the topocentric reference frame, expressed with Eq. \ref{eq7:posTopo}. For the remainder of this work, we express all position and velocity quantities in the topocentric reference frame unless otherwise specified. Given azimuth and elevation angle measurements, combined with range information, one can obtain a position estimate by algebraically manipulating Eq. \ref{eq8:AZ-EL}. 

\begin{figure}[h!]
	\centering\includegraphics[width=6in]{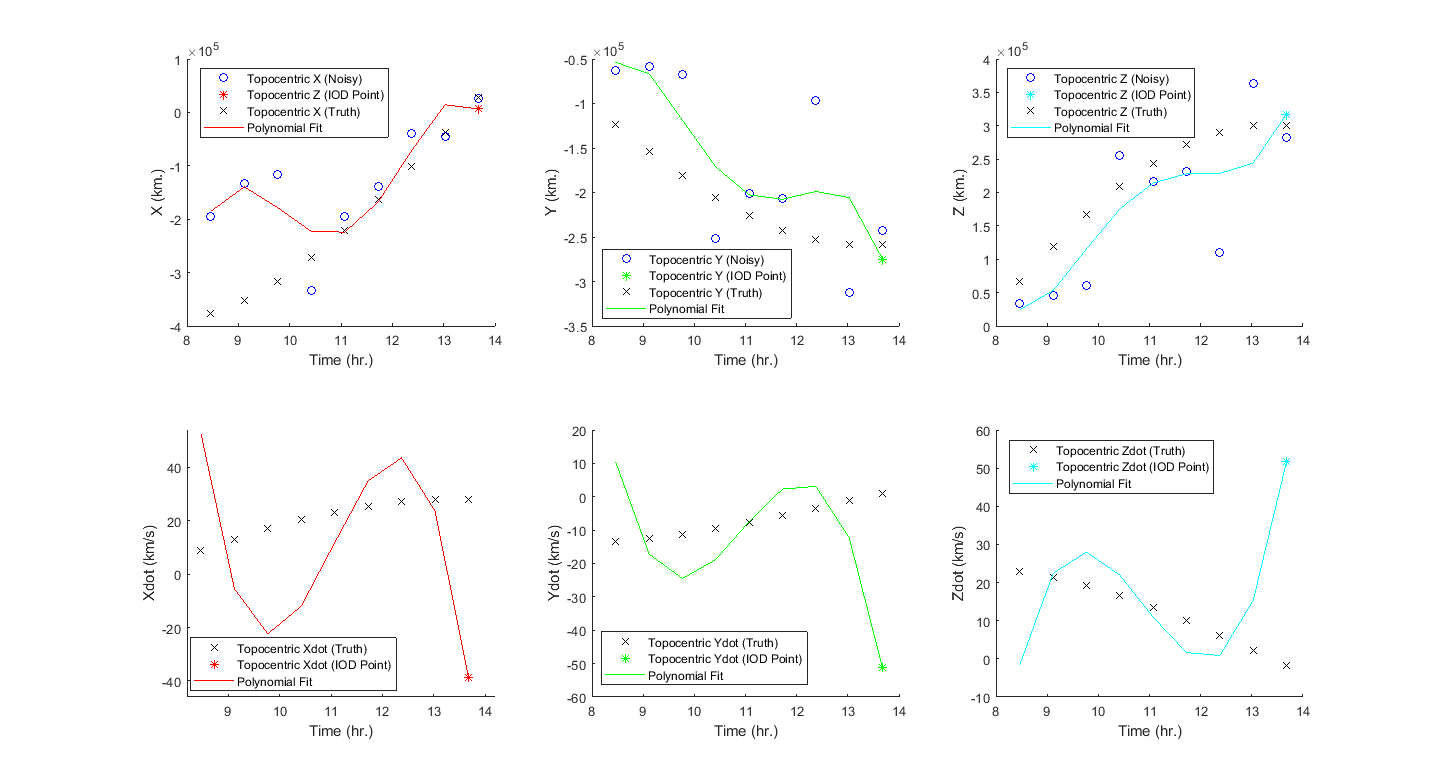}
	\caption{Kinematic polynomial fitting illustration over the first (50\%) of an RSO's pass, totaling 9 measurements (illustrated as blue circles) over roughly 5 hours. We fit fourth-order polynomials (in red, green, and cyan) as functions of $t$ through range, azimuth, and elevation measurements, converted to  $x$, $y$, and $z$ in the observer-centered topocentric reference frame using Eq. \ref{eq8:AZ-EL}. When we take the temporal derivatives for $x$, $y$, and $z$, an initial state estimate is formed at the last timestep in these plots (indicated by a star symbol). Because a single estimate will not accurately capture the target state, we simulate thousands of time-series sets of angular observations to obtain an initial state estimate. For comparison purposes, the target truth (denoted by a black cross) shows how closely our curve fitting approach matches the true state of the object over time.}
	\label{fig:1polyIOD}
\end{figure}

Since we are able to make multiple observations of an RSO in a single pass and note the times at which we make each observation, we can approximate the motion of a target during any fraction of the pass strictly as a function of time. We are then able to fit a polynomial spline curve through some fraction of the pass, and extract coefficients for the polynomial fit using a using maximum likelihood (ML) estimator. More specifically, we use the least-squares technique to obtain polynomial coefficients. We fit the same type of curve for each of $x(t)$, $y(t)$, and $z(t)$. With our coefficients, we simply take the time derivative of each curve or spline over the length of the interpolation interval to obtain estimates for the velocities $\mathbf{v}(t) = \dot{\mathbf{r}}(t) = [\dot{x}(t), \dot{y}(t), \dot{z}(t)]^T$. Since we know the time of the last observation through which we kinematically fit a curve, $t_f$, we may obtain a single initial state estimate $\mathbf{x}(t_f) = [\mathbf{r}(t_f)^{T}, \mathbf{v}(t_f)^{T}]^T$ for a given point of time. An illustration of the procedure for obtaining a single initial state estimate is given by Figure \ref{fig:1polyIOD}. 

\begin{figure}[h!]
	\centering\includegraphics[width=6in]{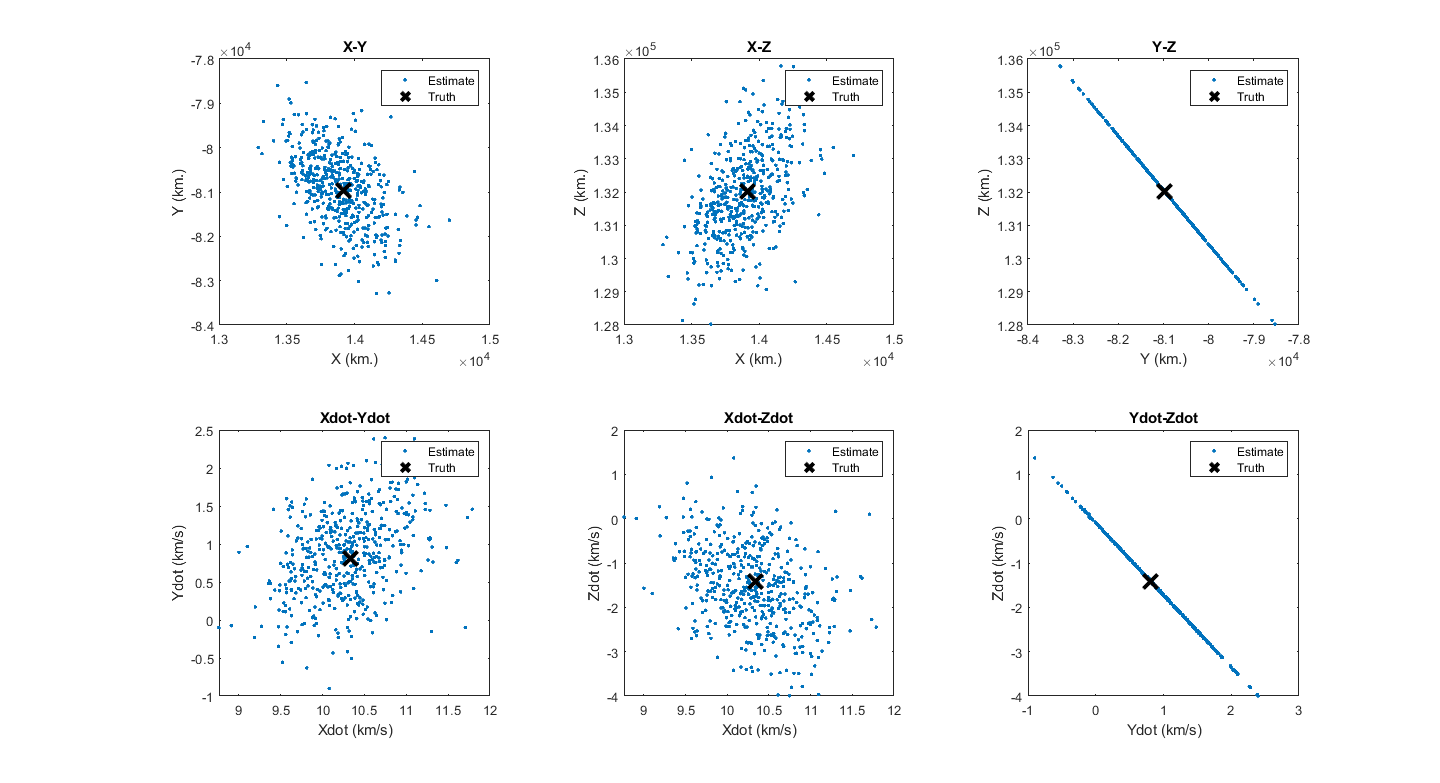}
	\caption{To obtain the initial state estimate in particle cloud form, we use the polynomial fitting method shown in Figure \ref{fig:1polyIOD} several hundreds or thousands of times, drawing each measurement and range information from the noise statistics or using \textit{a priori} knowledge of our target's orbit. Each point that would be indicated by a star from Figure \ref{fig:1polyIOD} would correspond to a single blue particle in this figure. To visually verify that our initial state estimate is consistent with the truth, it is clear that the target truth, given as a black cross, lies within the particle cloud -- abstracted into a probability density function (PDF) -- that makes a probabilistic initial state estimate.}
	\label{fig:2iodExample}
\end{figure}

Combining the facts that measurement noise for both azimuth and elevation angles is distributed as zero-mean Gaussian and that polynomial coefficients are obtained using least-squares optimization, a large number of initial state estimates is expected to fall both above and below the truth values. For this reason, it is important to utilize the polynomial fitting approach in this section using Monte Carlo methods. Measurements may be generated for each time step by bootstrapping the measurement noise statistics for thousands, even tens of thousands, of times. By the law of large numbers, the resulting initial state estimates will yield a large particle cloud within which the target truth is encapsulated, the volume of which may be expressed as a probability density function (PDF). A sample probabilistic initial state estimate, is provided in Figure \ref{fig:2iodExample}. The kinematic fitting IOD process is summarized in Algorithm \ref{alg:kf_iod}.

\begin{algorithm}
\caption{Kinematic Fitting Algorithm}\label{alg:kf_iod}
\begin{algorithmic}[1]
    \State Given $S$ time-varying $AZ$ and $EL$ measurements at time steps $t \in [t_i, \cdots, t_f]$ for a given pass.
    \For{$i = 1$ to $N$}
        \For{$s = 1$ to $S$}
            \State Sample $AZ(t)$ and $EL(t)$ from measurement noise statistics.
            \State Sample range information $\rho(t)$ from measured or inferred range statistics.
        \EndFor
        \State \parbox[t]{0.9\linewidth} {Translate all $[\rho(t), AZ(t), EL(t)]$ into the observer-centered topocentric position coordinates by algebraically manipulating Eq.~\ref{eq8:AZ-EL}.}
        \State \parbox[t]{0.9\linewidth} {Use an appropriate order polynomial or spline fitting method to obtain topocentric position estimates $[\overline{x}(t), \overline{y}(t), \overline{z}(t)]$ as functions of time.}
        \State \parbox[t]{0.9\linewidth} {Take temporal derivatives of each of $[\overline{x}(t), \overline{y}(t), \overline{z}(t)]$ to get $[\overline{\dot{x}}(t), \overline{\dot{y}}(t), \overline{\dot{z}}(t)]$.}
        \State \parbox[t]{0.9\linewidth} {Choose the full-state approximation at the final time step $[\overline{x}(t_f), \overline{y}(t_f), \overline{z}(t_f), \overline{\dot{x}}(t_f), \overline{\dot{y}}(t_f), \overline{\dot{z}}(t_f)]$. This becomes a single initial state estimate of your probabilistic initial state estimate.}
    \EndFor
\end{algorithmic}
\end{algorithm}

The shape and size of the initial state estimate will depend upon several factors, such as measurement noise, the type of range information used, and the number of measurements through which one obtains an initial state estimate. Generally, if one fits polynomials through a smaller number of observations, the size of the initial state estimate becomes smaller. However, this initial state estimate tends to become overconfident and a nonlinear filter breaks down with subsequent update steps. With sufficient trial and error, it may be observed that a large number of consecutive observations -- typically 8-12 -- fitted with a fourth-order polynomial tends to sufficiently avoid an overconfident initial state estimate. 

The key requirement for our IOD method is a time series of initial position estimates. Although optical telescopic data provide sufficiently precise and accurate angular information about the location of a target relative to an observer, range information can be difficult to infer due to the vast distances within cislunar space. To compensate for this difficulty in obtaining range information, we can make one of two reasonable assumptions: 1) Assume a high measurement noise for the range $\rho$, or 2) Assume that our target lies within the bounds of cislunar space. For the first assumption, we assume that the range measurement noise is roughly some non-insignificant percent of the true target range. For example, if a target is roughly 400000 km. away from our observer, we assume that the range "measurement" noise of our target is distributed as zero-mean Gaussian with a standard deviation of 5-10\% of the true range, or between 20000 to 40000 km. For the second assumption, we assume that we have no information about the target range except that it lies within the bounds of cislunar space. This allows us to generate a uniform PDF $\rho \sim \mathcal{U}[84328, 550000]$ km. to model maximum range uncertainty. With the law of large numbers and sufficient pruning of initial state estimates that lie outside of the cislunar domain, we develop a large but consistent initial state estimate. In the proceeding section, we shall discuss the utility of each of these assumptions on developing an initial state estimate and filter performance.

Regardless of which assumption we use, our initial state estimate is typically large; one standard deviation in the position space tends to be tens of thousands of kilometers while one standard deviation in the velocity space tends to be tens of kilometers per second! However, a good nonlinear filter can dramatically reduce the size of this initial estimate with subsequent observations. Next, we describe a filter capable of handling the nonlinearities associated with both the dynamics of cislunar space as well as the angular observations of an optical telescope. 

\subsection{Relationship to Gauss' method for IOD in Keplerian Orbits}\label{subsec:relation}

Gauss' method for orbit determination in Keplerian orbits uses three angles measurements, say $\bar{\alpha} = (\alpha_1, \alpha_2, \alpha_3)$ and $\bar{\delta} = (\delta_1, \delta_2, \delta_3)$, to find the associated range to the target $\bar{\rho} = (\rho_1,\rho_2,\rho_3)$. In order to accomplish this: 1) the method uses a Taylor series in time for the position vector, 2) uses the two body dynamics to express the position vector at any time as a linear combination of the position and velocity vectors at some initial time, and 3) uses the fact that the second position vector can be expressed as a linear combination of the first and last position vectors because Keplerian motion is planar: $\bar{r}_2 = c_1 \bar{r}_1 + c_3 \bar{r}_3$.\\
These three steps allow us to formulate a set of three nonlinear equations in the components of the unknown range vector $\bar{\rho}$ that we can then solve to obtain the position vectors, and step 2) is then used to find the velocity vector at the second time step. The critical assumption is in step 2) where the two body dynamics allows us to write the position at some time as a linear combination of the position and velocity vectors at some initial time: 
\begin{align}\label{Eq.fg}
\bar{r}(t) = f(\rho_0,t-t_0)\bar{r}(t_0) + g(\rho_0,t-t_0)\dot{\bar{r}}(t_0),
\end{align}
where $f,g$ are known functions of $\rho_0$ and the time difference $(t-t_0)$. The key element to note is that even Gauss' method uses a time interpolation through the measurements like in our approach. An initial state cloud can be sampled for Gauss' method similar to our approach by sampling angles $(\bar{\alpha}^i, \bar{\delta}^i)$ perturbing the measurements with the sensor noise and finding the corresponding ranges $\bar{\rho}^i$ using the procedure above. The key is that in Gauss' method, due to the Keplerian motion the range vector $\bar{\rho}^i$ is a function of the sampled angles $(\bar{\alpha}^i, \bar{\delta}^i)$.\\

Unfortunately, in the three body problem, step 2) is no longer valid and we cannot write the position vector as in equation \eqref{Eq.fg}. Therefore, it is not possible to solve for the range vector $\bar{\rho}$ as a function of the angles $(\bar{\alpha}, \bar{\delta})$ as above. Instead, we sample the range vectors $\bar{\rho}^i$ from some underlying distribution, that models some a priori information such as that we believe the object is in cislunar space, independently of the angles measurements. The resulting position vectors are then fitted in time using a polynomial and the polynomial is then differentiated to get the velocity vector. The drawback is that the sampled ranges may not be accurate and the resulting initial state cloud may have large uncertainty. However, our philosophy is to get a consistent initial state pdf estimate making minimal assumptions regarding the orbit and then use subsequent measurements allied with a suitable tracking approach that utlizes the three body dynamics to get an accurate estimate of the state over time as we show in the following development. We note that a good filter should not need an accurate initial state pdf, and in fact, the subsequent measurements dominate the filtered pdf since initial conditions should be forgotten exponentially by a good filter \cite{raihan2018}. \\
A final note is that in Gauss' method, albeit it seems that $\bar{\rho}^i$ is a function of the angles $(\bar{\alpha}^i, \bar{\delta}^i)$, it involves the solution of a nonlinear set of equations whose solution critically depends on the initial guess at these ranges. Thus, implicitly, even in Gauss' method, if this initial guess is made in a random fashion, it is equivalent to a sampling for the range vector as in our approach.

\subsection{Orbit Determination}\label{subsec:OD}

As discussed in Section \ref{sec:2prelims}, our dynamics and measurement models are both nonlinear. On top of that, kinematically fitting polynomials over sequences of data will render our probabilistic initial state estimate to slightly deviate from a Gaussian PDF. This deviation is considerable in the measurement space. For this reason, Extended Kalman Filters (EKF), Unscented Kalman Filters (UKF), and particle filters can fail to accurately capture the non-Gaussianity of the initial state estimate and proceeding estimates. Instead of fitting our particle-based estimates to a Gaussian distribution, we model all \textit{a priori} estimates with a Gaussian Mixture Model (GMM). Using a proper filtering framework will allow us to control the non-Gaussianity of our state estimates over time.

Starting with an initial state estimate such as that produced by Figure \ref{fig:2iodExample}, orbit determination (OD) for subsequent timesteps shall utilize the recursive Particle Gaussian Mixter (PGM) Filter. The PGM Filter was first developed to address two key issues in GMM based filters: 1) inflexibility of the component numbers and weights while propagating the estimate, and 2) the curse of dimensionality in the form of particle depletion \cite{raihan2016, raihan2018} The development of this filter was based on a successful implementation of a hybrid UKF and particle filter based approach for tracking objects in the LEO regime.\cite{raihanukfpf2018} 

In this subsection, we demonstrate the PGM Filter for a single timestep. The PGM Filter consists of four key recursive steps: propagation, clustering, measurement update (or simply update for short), and resampling. Starting with a particle-based initial estimate, we propagate each particle with our nonlinear prediction model until we are able to observe the target. Process noise is added to each particle at this step only if we are either using a filter which tends to perform with overconfidence (such as an EKF or GMM-based EKF) or if our truth dynamics are expected to deviate sufficiently from the underlying dynamics model. Once we are able to obtain an observation, we invoke the clustering and update steps. 

\begin{figure}[thpb] 
      \centering
      \begin{subfigure}{\textwidth}
            \centering
            \includegraphics[width=\linewidth]{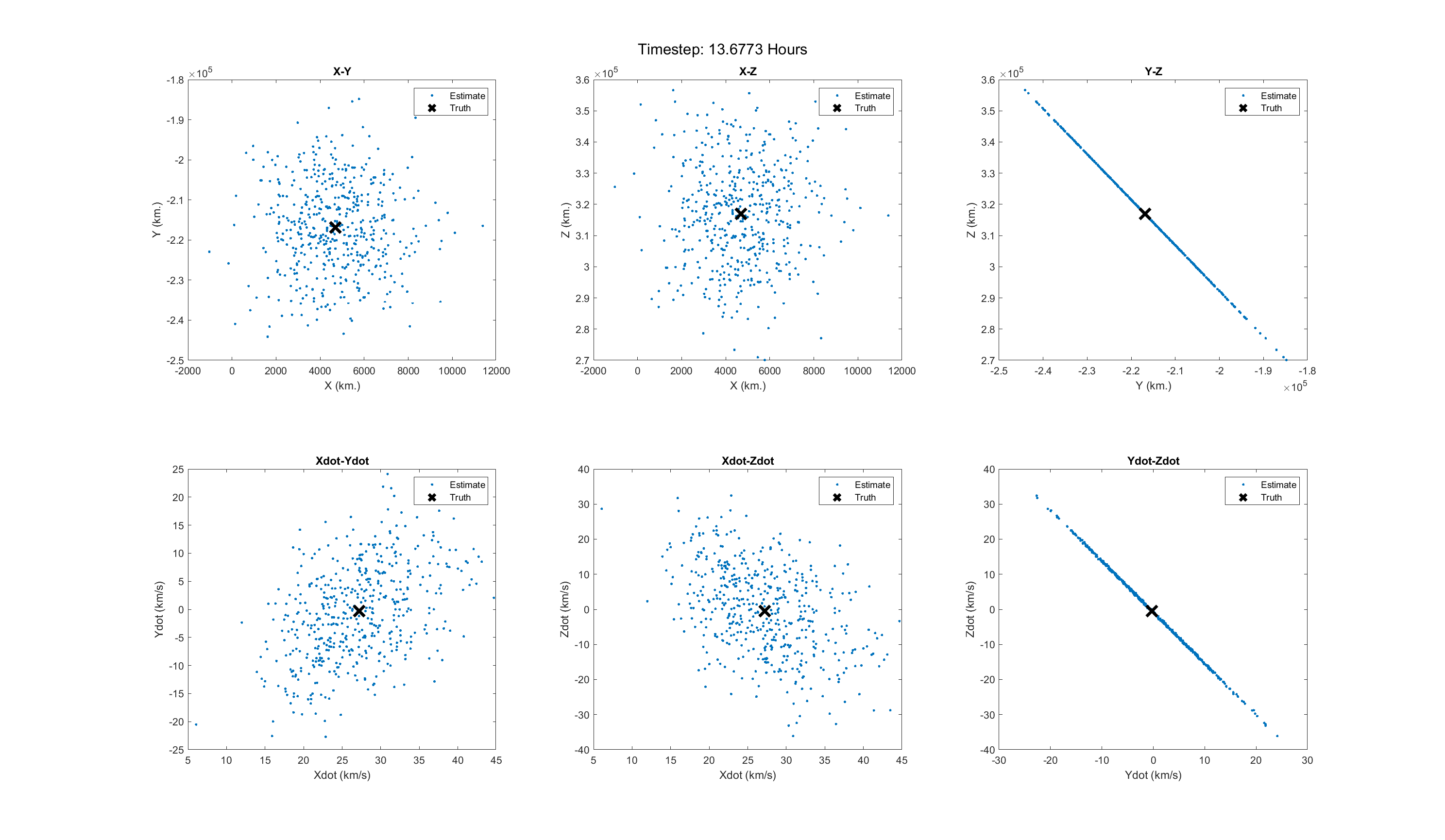}
            \caption{Initial State Estimate (IOD)} \label{fig:3aiodWilliams}
      \end{subfigure}
      \vspace{0.5cm} % optional vertical spacing
      \begin{subfigure}{\textwidth}
            \centering
            \includegraphics[width=\linewidth]{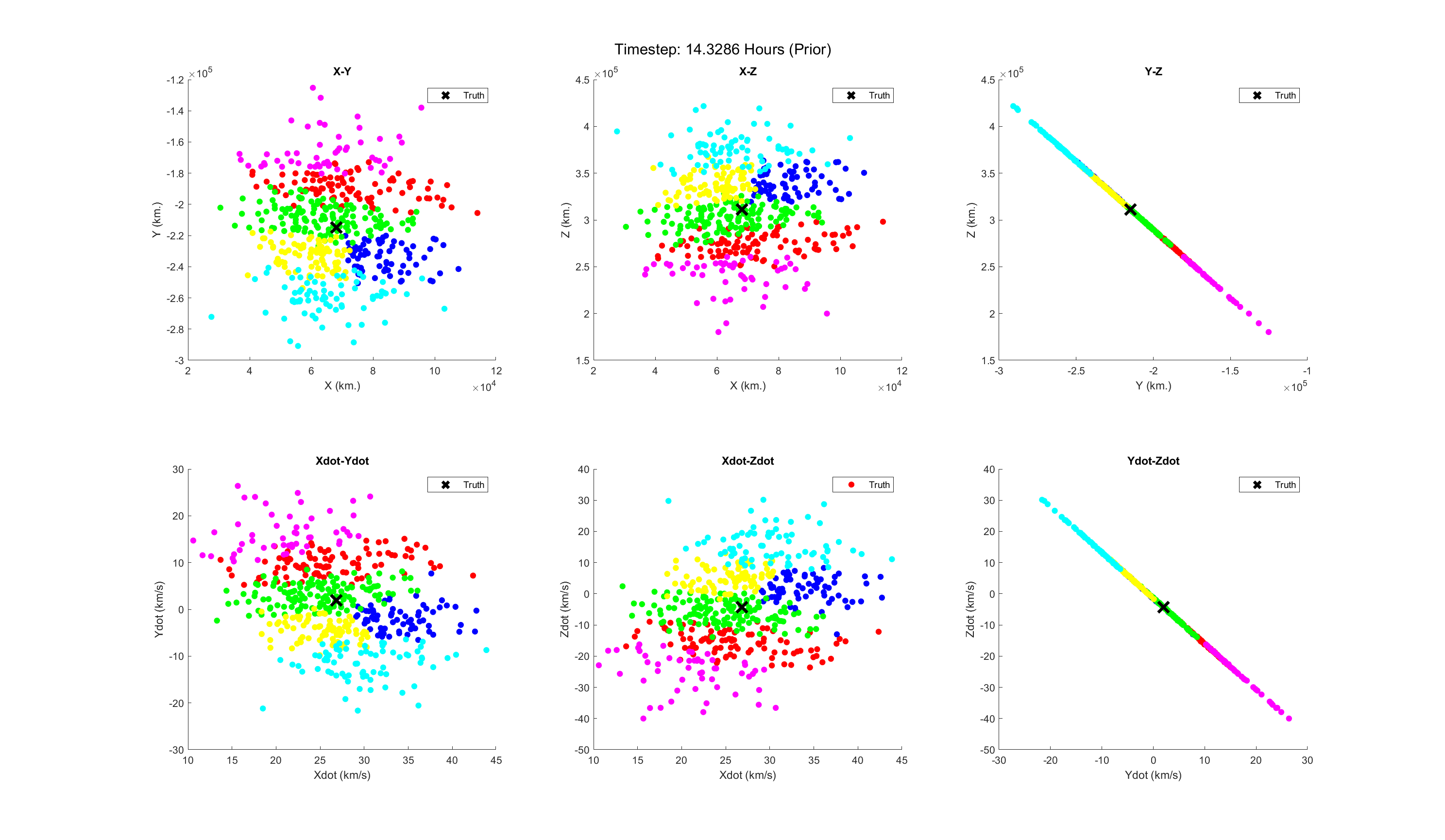}
            \caption{Propagation and clustering steps}\label{fig:3bAprioriEx_Williams}
      \end{subfigure}

      \caption{Using the polynomial fitting method shown in Figure \ref{fig:1polyIOD} several hundreds or thousands of times, we develop an initial state estimate. The choice of trajectory for the target (denoted by a black X) is one which is propagated by a CR3BP patch point as described by \textit{Williams, et al. 2017}~\cite{williams2017}. Using \textit{k-means} clustering, we partition our propagated estimate into six clusters, representing \textit{a priori} GMM components. Based on the number of particles in each component, the cluster means, and cluster covariances, an \textit{a priori} PDF is generated for our orbit.}
    \label{fig:3williams2017iod_cluster}
\end{figure}

We can use any clustering algorithm $\mathcal{C}$ to distribute the particles of our initial state estimate into $M^{+}(n-1)$ clusters. By taking the mean and covariance of each of the clusters, we create GMM components for the \textit{a priori} estimate. For our article, we utilize the \textit{k-means} algorithm to segregate particles of our \textit{a priori} ensemble into clusters recursively.\cite{lloyd1982} To demonstrate our propagation and clustering steps, we begin with an initial state estimate derived from our IOD method. This initial state estimate is expressed in particle form in Figure \ref{fig:3williams2017iod_cluster}, and is roughly centered at the target truth. This particle ensemble is propagated to yield the \textit{a priori} estimate for the subsequent time step. Due to the size of the initial state estimate, we utilize six clusters.

\begin{figure}[h!]
	\centering\includegraphics[width=6in]{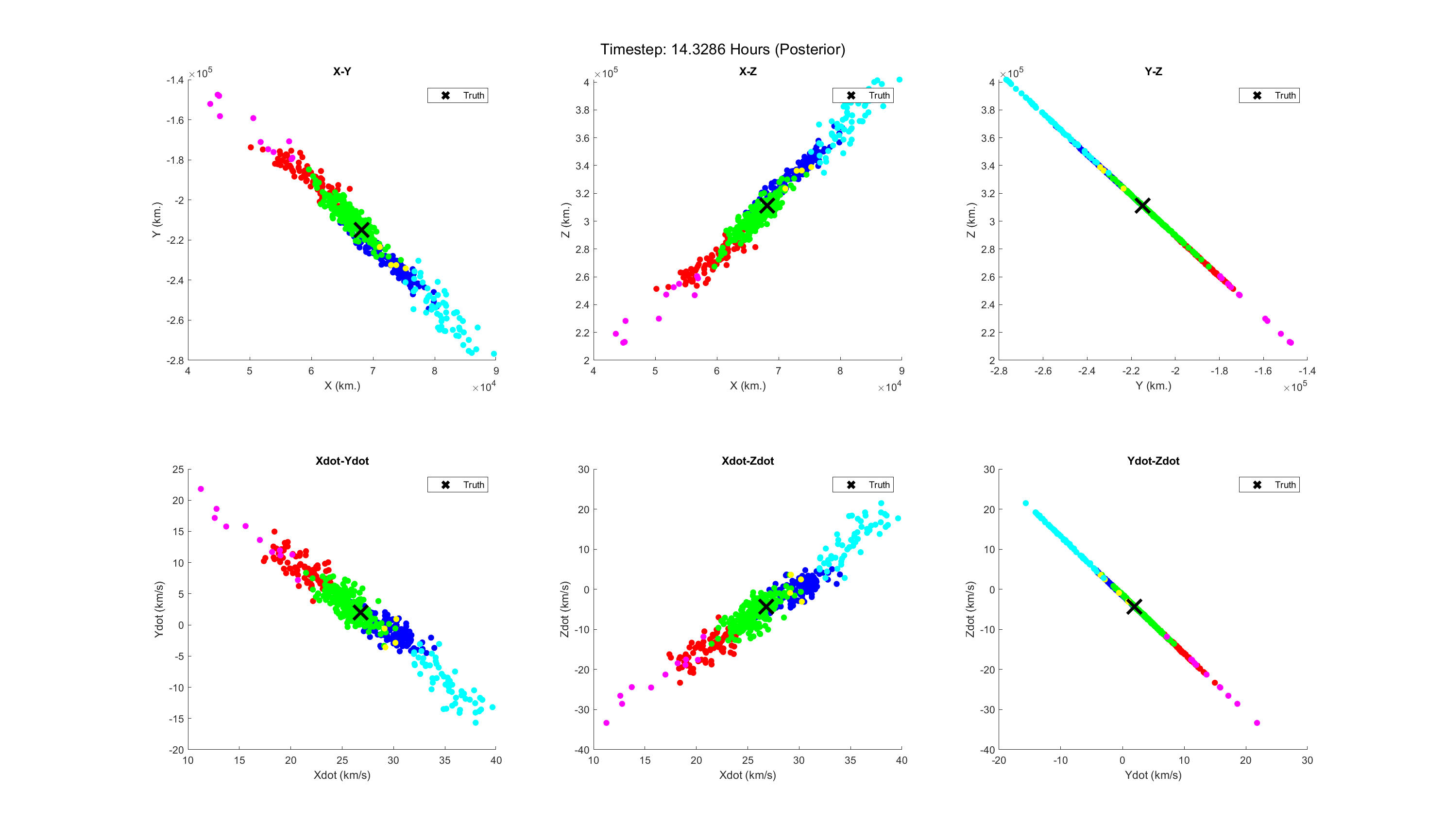}
	\caption{Component-wise update and resampling steps to the estimate shown in Figure \ref{fig:3bAprioriEx_Williams}. As expected, the cluster defined in green is expected to retain a large fraction of the weight update. Due to low measurement likelihoods, the purple and yellow clusters will vanish.}
	\label{fig:4posteriorWilliams}
\end{figure}

At this point, we fuse our angular observations with our \textit{a priori} GMM-based estimate using an Ensemble Kalman Filter (EnKF) update. Since our \textit{a priori} ensemble is expressed as a GMM, we update each component prior mean $\mathbf{\mu}_i^{-}(n)$ and covariance $\mathbf{P}_i^{-}(n)$ of each ensemble with the following equations:
\begin{subequations} \label{eq7:enkfUpdate}
    \begin{align}
        \mathbf{\mu}_i^{+}(n) = \mathbf{\mu}_i^{-}(n) + \mathbf{P}_{i,zx}^{T}(n)\mathbf{P}_{i,zz}^{-1}(n)(\mathbf{z}(n) - \mathbb{E}_i[\mathbf{h}(\mathbf{X})]) \label{eq7a:muUpdate} \\
         \mathbf{P}_i^{+}(n) = \mathbf{P}_i^{-}(n) - \mathbf{P}_{i,zx}^{T}(n)\mathbf{P}_{i,zz}^{-1}(n)\mathbf{P}_{i,zx}(n) \label{eq7b:covUpdate}
    \end{align}
\end{subequations}
where
\begin{subequations} \label{eq8:enkfUpdateClarifs}
    \begin{align}
        \mathbf{P}_{i,zx}(n) = \mathbb{E}_i[(\mathbf{h}(\mathbf{X}) -  \mathbb{E}_i[\mathbf{h}(\mathbf{X})])(\mathbf{X} - \mathbb{E}_i[\mathbf{X}])^{T}],\label{eq8a:Pzx} \\
         \mathbf{P}_{i,zx}(n) = \mathbb{E}_i[(\mathbf{h}(\mathbf{X}) -  \mathbb{E}_i[\mathbf{h}(\mathbf{X})])(\mathbf{h}(\mathbf{X}) -  \mathbb{E}_i[\mathbf{h}(\mathbf{X})])^{T}].\label{eq8b:Pzz} 
    \end{align}
\end{subequations}
In the above sets of equations, $\mathbf{X}$ defines the set of particles in our estimate, and $\mathbf{h}$ defines the measurement model (vectorized from Eq. \ref{eq7:posTopo}). For the component weight update, each component's \textit{a priori} weight is updated based on the likelihood that each cluster (in the measurement space) explains the underlying observation. At the end of the update step, particles are resampled from the new GMM distribution, and the filter continues to be applied recursively. Using Eqs. \ref{eq7:enkfUpdate} - \ref{eq8:enkfUpdateClarifs}, we show the update step to the estimate in Figure \ref{fig:4posteriorWilliams} and the resampled particles. A summary of the PGM Filter is outlined in Algorithm \ref{alg:pgmf}.\cite{raihan2018}

\begin{algorithm}
\caption{Particle Gaussian Mixture Filter Algorithm}\label{alg:pgmf}
\begin{algorithmic}
    \State Given $\pi_0 (x) = \sum_{i=1}^{M(0)} \omega_i (0) p_g(x; \mu_i (0), P_i (0))$, transition density kernel $p(x'|x)$, $n = 1$.
    \begin{enumerate}
        \State Sample $N_p$ particles from $\pi_{n-1}$ and the transition density kernel $p_n (x'|x)$ as follows: \label{alg:repeat}
        \begin{itemize}
            \item Sample $X^{(i)^\prime}$ from $\pi_{n-1} (.)$.
            \item Sample $X^{(i)}$ from $p(.|X^{(i)^\prime})$.
        \end{itemize}
        \State Use a clustering algorithm $\mathcal{C}$ to cluster the set of particles $X^{(i)}$ into $M^{-} (n)$ Gaussian clusters with weights, mean, and covariance given by $\{\omega_i^{-} (n), \mu_i^{-} (n), P_i^{-} (n)\}$.
        \State Update the mixture weights and mixture means and covariances to $\{\omega_i^{+} (n), \mu_i^{+} (n), P_i^{+} (n)\}$, given the observation $z_n$, utilizing the Kalman update. 
        \State $n = n + 1$. Go to Step \ref{alg:repeat}.
    \end{enumerate}
\end{algorithmic}\end{algorithm}

The PGM Filter may be tuned by three variables: the number of particles, the number of clusters, and the clustering strategy. In effect, most of the tunability of the PGM Filter lies at the clustering step. The number of particles and the number of clusters will generally depend upon the shape and size of the PDF describing the target. For example, after the propagation step immediately following IOD, we would need to utilize a large number of particles (typically in the thousands or even tens of thousands) and a sufficiently large number of clusters (typically between 4-8) in order to accurately represent the size and non-Gaussianity of the predicted estimate. The exact number of particles or clusters utilized for this filter will vary with time and non-Gaussianity of the state estimate. Although the particle-based representation of state estimates, EnKF-based update step, and resampling step make a filter robust to highly nonlinear dynamics and highly non-Gaussian PDFs, the success of the PGM Filter is highly dependent upon an effective clustering strategy. While \textit{k-means} clustering is an effective strategy for our application, more sophisticated clustering strategies can be developed and is a subject of ongoing research.

\section{Results and Discussion}\label{sec:4results}

Within this section, we shall discuss and demonstrate the utility of our combined, minimal-assumption IOD-OD framework. As mentioned in Section \ref{subsec:IOD}, we can generate initial state estimates through inferring range information by accepting large range measurement errors or by drawing range information from a uniform PDF whose bounds encompass the cislunar domain. In this section, our examples and discussion shall focus on both IOD generation techniques. We shall first demonstrate the utility of our IOD-OD framework upon three different orbits/trajectories. Then, we shall discuss the performance of the PGM filter with respect to other common filters within the cislunar domain. After that, we shall compare the effects of differing levels of initial range and/or velocity information upon generating initial state estimates and long-term behavior of our IOD-OD framework. We shall conclude the section by discussing limitations to these minimal-assumption IOD and target tracking approaches.

\subsection{Important Noise Characteristics and Metrics}\label{subsec:metrics}

After the IOD step, we utilize angles-only measurements for this work. Assuming measurement noise to be additive and distributed as zero-mean and Gaussian, we define our noise covariance matrix $\mathbf{R}$ as 

\begin{equation}\label{eq9:OD_R}
    \mathbf{R} =
    \begin{bmatrix}
        (1.5 \space \text{arcsec})^2 & 0\\
        0 & (1.5 \space \text{arcsec})^2
    \end{bmatrix}
\end{equation}
due to the high, arcsec-level precision of modern telescopic sensors.\cite{mishra2024} For our measurement model, we assume that our observer is a lidar-type sensor based in College Station, Texas with high angular precision and the capability to measure distances in cislunar space with low precision. Due to the ranges associated with the cislunar domain relative to those up to the GEO limit, we assume a very high range "measurement" noise, expressed as some fraction $\alpha$ of the true range, denoted as $\rho^{*}$. During the IOD stage, the measurement noise covariance matrix may be expanded to 
\begin{equation}\label{eq10:iod_R}
    \mathbf{R} =
    \begin{bmatrix}
        (\alpha \rho^{*})^2 & 0 & 0\\
        0 & (1.5 \space \text{arcsec})^2 & 0\\
        0 & 0 & (1.5 \space \text{arcsec})^2
    \end{bmatrix},
\end{equation}
where $\alpha$ is typically between 0.05-0.10. Although this implies a concerning single standard deviation error of 20000 km. for an object close to the Moon, we shall demonstrate later in this section that such a high range error does not pose an issue for the IOD-OD framework in the short or long term.

To analyze the performance and dilution of precision of the PGM Filter over time, we shall utilize an entropy metric $\mathbf{H}(k)$, given by 
\begin{equation}\label{eq5:entropy}
    H(k) = \mathbb{E}[- \log{p_{k}(\mathbf{x})}] \approx -\frac{1}{N} \sum_{i=1}^{N} \log{p_k(\mathbf{x})},
\end{equation}
where $p_k(\mathbf{x})$ refers to particle estimate or PDF at time step $k$, often expressed as a GMM. Since we use a particle-based representation of our GMM PDF, we utilize an $N$-particle Monte Carlo approximation of our entropy, given on the right hand side of Eq. \ref{eq5:entropy}. For a UKF or Ensemble Kalman Filter (EnKF), $p_k(\mathbf{x})$ is a Gaussian PDF for all $k$.

\subsection{Example 1: 9:2 Resonant Near-Rectilinear Halo Orbit}\label{subsec:ex1}

\begin{figure}[h!]
	\centering\includegraphics[width=6in]{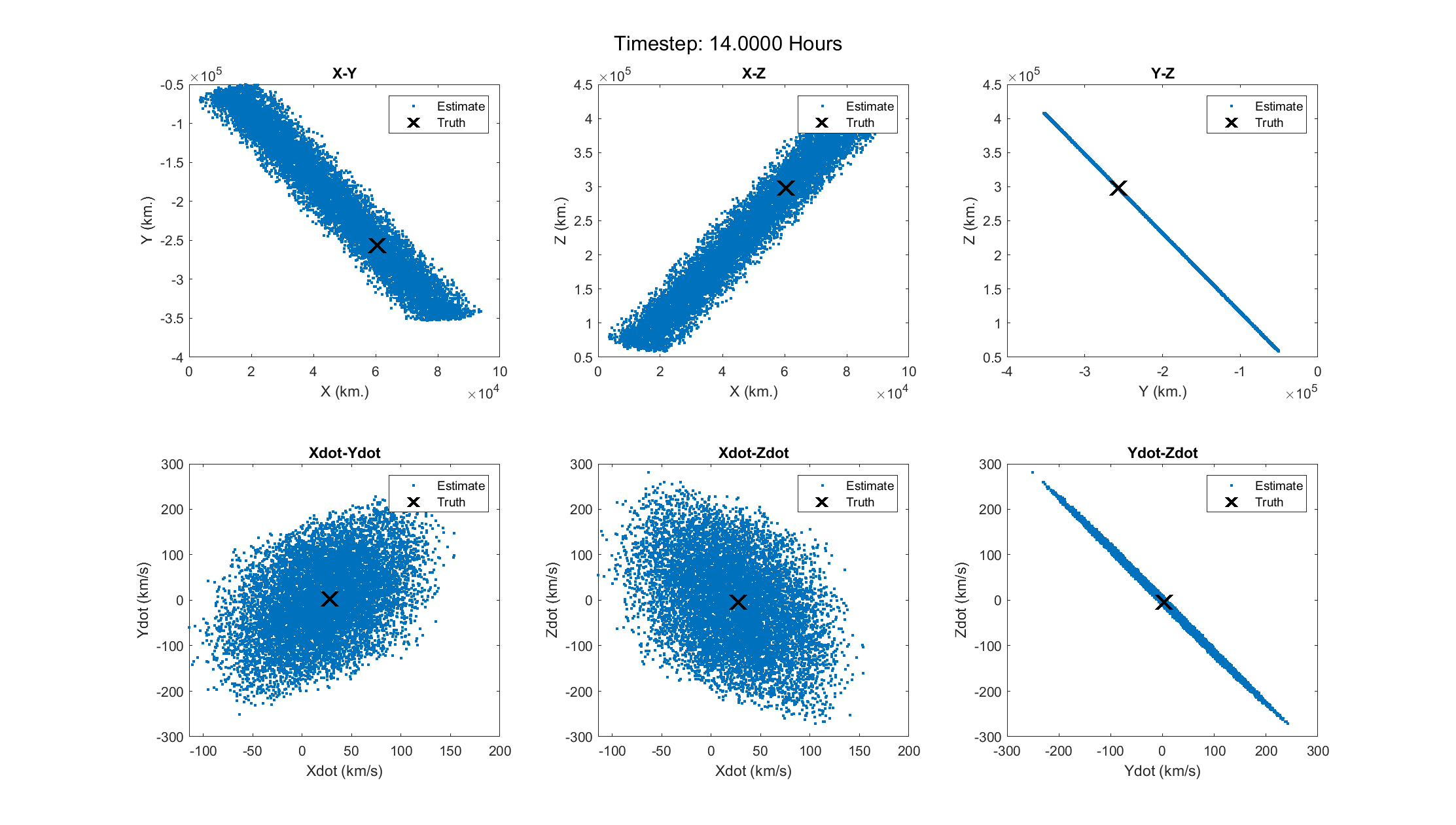}
	\caption{Initial state estimate derived from kinematically fitting polynomials through several series of possible measurements whose range information is derived from a large uniform PDF. Particles in the position space are pruned if they lie outside of the bounds of cislunar space. Although the resulting PDF is large and very uncertain in the range direction, it is localized in terms of AZ and EL, and we may utilize the PGM Filter to reduce the uncertainty.}
	\label{fig:5iodExample1}
\end{figure}

NASA's Gateway program has chosen a 9:2 resonant near-rectilinear halo orbit (NRHO) as its base of operations. NASA chose this orbit due to low $\Delta v$ requirements for the Orion spacecraft and other potential targets operating within the vicinity and communication coverage about the Moon's south pole.\cite{2019NASA} Due to the relevance and significance of this orbit for upcoming lunar missions, we demonstrate the utility of our combined IOD-OD framework for a target in this orbit. We start by generating thousands of time-series sets of admissible measurements spaced roughly 40 minutes to one hour apart. For each measurement of each measurement set, we randomly sample a range value $\rho$ (in km.) from the aforementioned uniform distribution $\rho \sim \mathcal{U}[84328, 550000]$ km. After converting each generated $\{\rho(t), AZ(t), EL(t)\}$ set into the topocentric reference frame with Eq. \ref{eq7:posTopo}, we kinematically fit a fourth-order polynomial through the first 50\% of measurements of the first pass for each measurement set and take the derivative of each polynomial to obtain an initial state estimate at the final timestep through which we fitted each polynomial and repeat this procedure over thousands of measurement sets. In other words, we utilize a PAR-like procedure to generate a very large PDF for the range and kinematically fit polynomials for our initial state estimates accordingly. The resulting initial state estimate is given by Figure \ref{fig:5iodExample1}.

\begin{figure}[thpb] 
      \centering
      \begin{subfigure}{\textwidth}
            \centering
            \includegraphics[width=\linewidth]{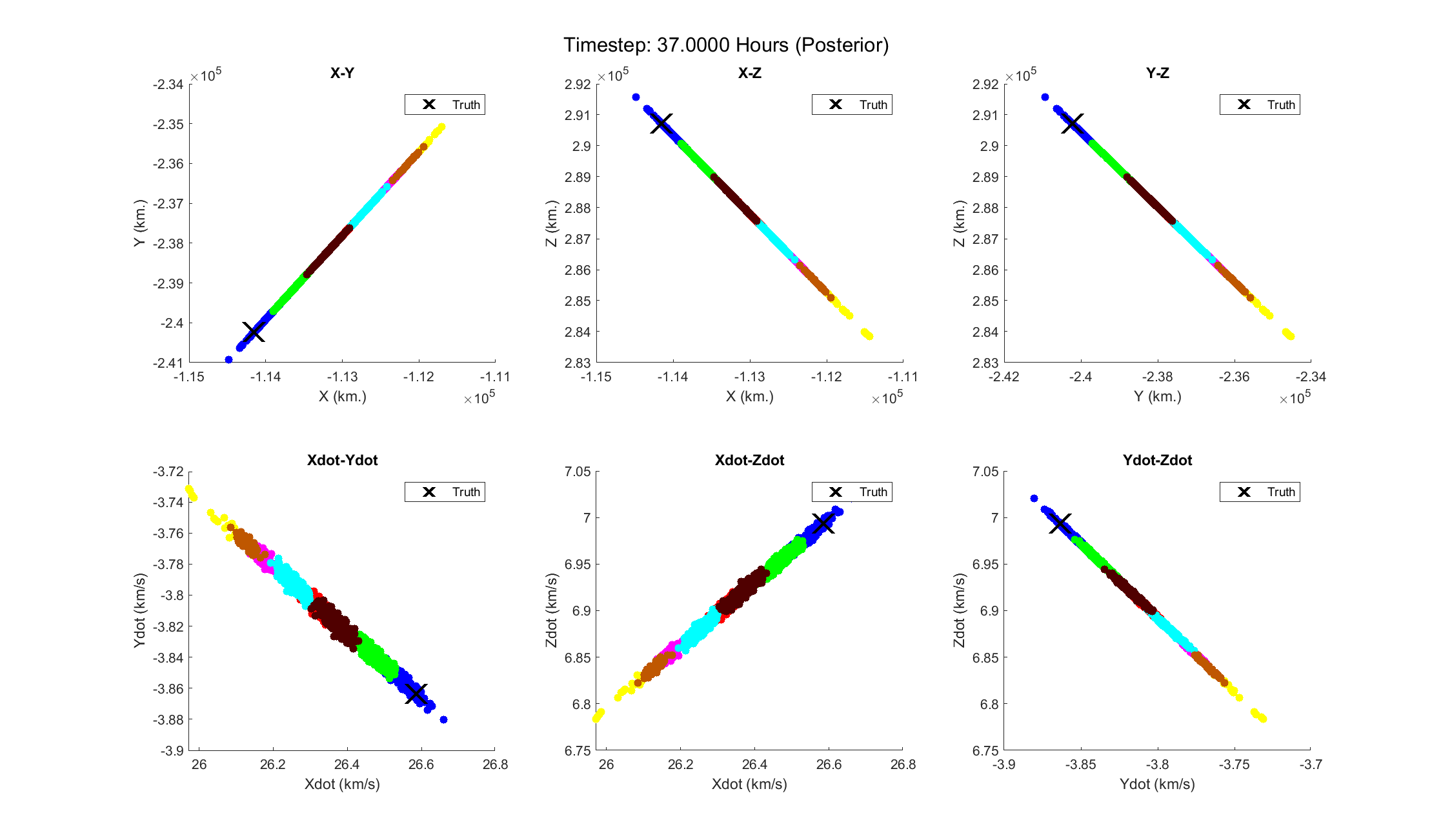}
            \caption{State Estimate at End of 2nd Pass} \label{fig:6a_endOf2nd}
      \end{subfigure}
      \vspace{0.5cm} % optional vertical spacing
      \begin{subfigure}{\textwidth}
            \centering
            \includegraphics[width=\linewidth]{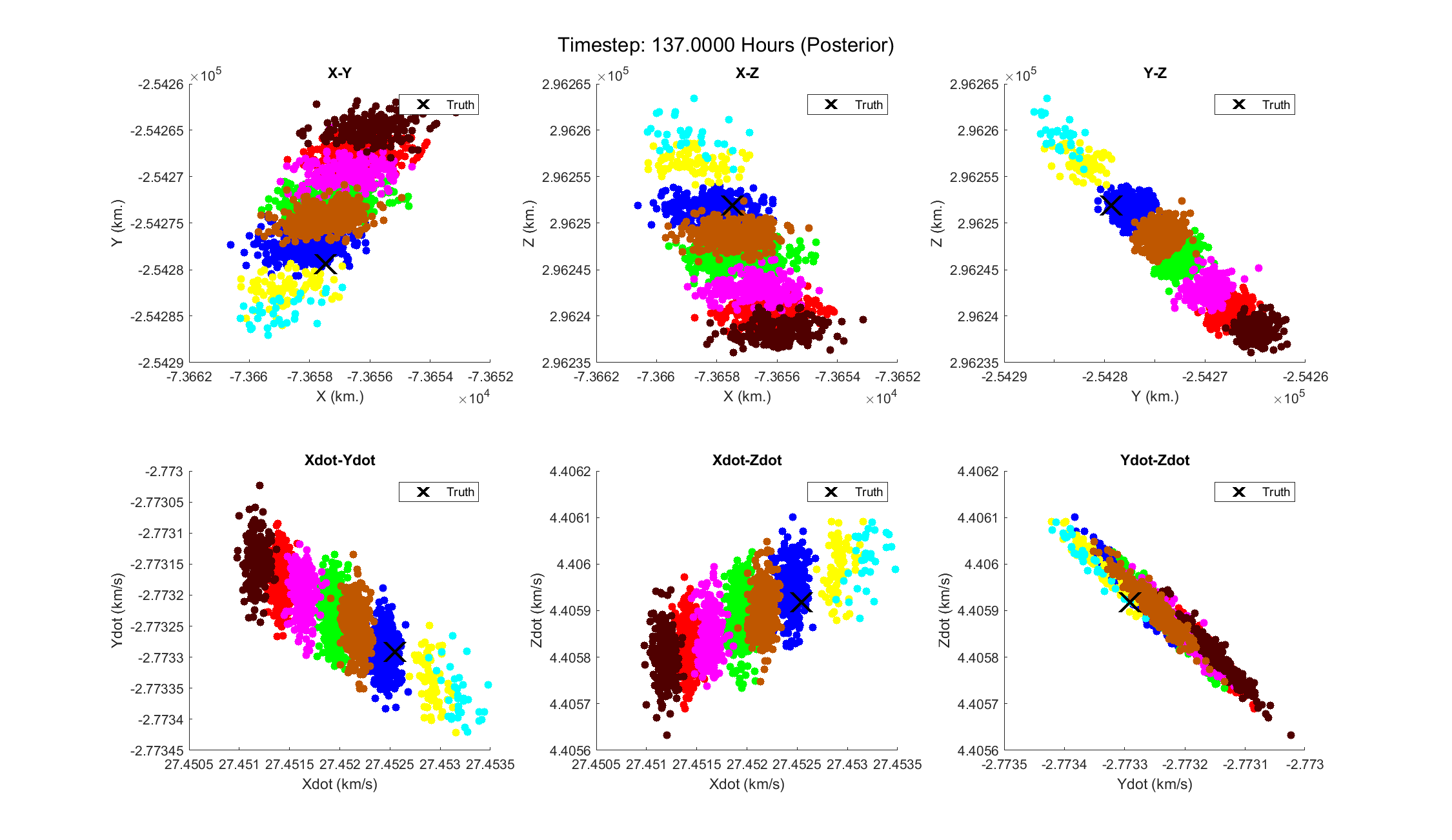}
            \caption{State Estimate at End of 6th Pass}\label{fig:6b_endOf6th}
      \end{subfigure}

      \caption{State estimates for a target in the 9:2 resonant NRHO. Standard deviations in the position estimate for each coordinate go from 1000-1300 km. in (a) down to 1.5-7.5 km. in (b). Similarly, standard deviations in the velocity estimate for each coordinate are reduced from 1.4-5.6 km/s in (a) to 18-45 cm/s in (b). This translates to over a 120-fold improvement in position accuracy and a ten-thousand fold improvement in the velocity accuracy!}
    \label{fig:6sample9-2results}
\end{figure}

Beyond the first pass, the observer is tasked to only take five random but consecutively spaced measurements for each subsequent pass. This tasking scenario was chosen to better mimic sensor tasking in reality. By utilizing our angles-only measurement model, we can drastically reduce the uncertainty of our target state estimate with each pass. Sample results for this orbit at the end of the second and sixth passes are provided in Figure \ref{fig:6sample9-2results}.

\begin{figure}[h!]
	\centering\includegraphics[width=8cm, height=6cm]{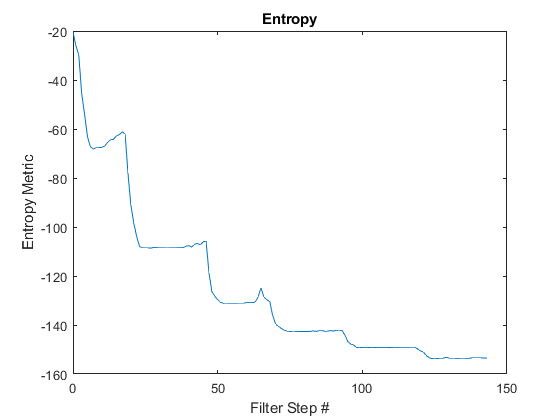}
	\caption{Entropy of the target state estimate over roughly 150 time steps corresponding to one full orbital period around the Moon (i.e. 6.25 Earth days). Between each pass, the entropy is flat or slightly increasing due to nonlinearity of dynamics and growing non-Gaussianity of the particle estimates. During the five time steps for each pass during which we obtain observations, entropy drops dramatically for the first 2-3 passes before flattening.}
	\label{fig:7ex1_entropy}
\end{figure}

To quantitatively analyze the dilution of precision for the state estimate, we use the entropy metric $E(k)$ defined by Eq. \ref{eq5:entropy}. The results are given by Figure \ref{fig:7ex1_entropy}.

For Sections \ref{subsec:ex2} and \ref{subsec:ex3}, we shall demonstrate our IOD/OD framework upon more chaotic orbits and trajectories.

\subsection{Example 2: Target Passing Through the $L_2$ Lagrange Point}\label{subsec:ex2}

The CR3BP Lagrange points along the Earth-Moon axis (i.e. $L_1$, $L_2$, and $L_3$) have long interested members of the astrodynamics community due to their unstable equilibrium property. The volume around these Lagrange points allows cislunar spacecraft to more efficiently execute orbit plane changes and transfers. The $L_2$ Lagrange point in particular has intrigued scientists because of the myriad orbit families and possible bifurcations resulting from minuscule position perturbations in the synodic $x$-direction and minuscule velocity perturbations in the synodic $y$-direction.\cite{grebow2006, zimovan2017, zimovanspreen2020} Indeed, the 9:2 NRHO orbit from our first example comes from a family of periodic orbits about this Lagrange point. 

A simple simulation that involves drawing several particles from a reasonably small Gaussian distribution about the $L_2$ libration point and propagating them for several hours results in a chaotically warped PDF whose Gaussianity breaks down severely. Due to the CR3BP's sensitivity to initial conditions, some of the worst Gaussianity breakdowns occur at state estimates centered near or at this libration point. For this example, we demonstrate how our IOD-OD framework controls the non-Gaussian behavior of a target estimate that is propagated past this libration point, demonstrating our filter's ability to provide reasonable state estimates in a chaotic dynamical system with limited information.

For this example, we utilize the same IOD method as the previous example. However, for OD, we task our sensor differently. For the first 33 hours, we attempt to take angular measurements of our target roughly once every 40 minutes. This length and frequency of measurements allows us to sufficiently reduce the uncertainty of our initial state estimate such that, after sufficient propagation of this new state estimate, we observe a sufficiently warped PDF. We let a 10-day sensor shutoff happen subsequently, during which we do not take any measurements. Once the sensor is capable of taking measurements again, we take measurements only once every 16 hours. We choose these parameters for our example to illustrate the utility of the PGM Filter for long-term sensor shutoffs.

\begin{figure}[h!]
	\centering\includegraphics[width=6in]{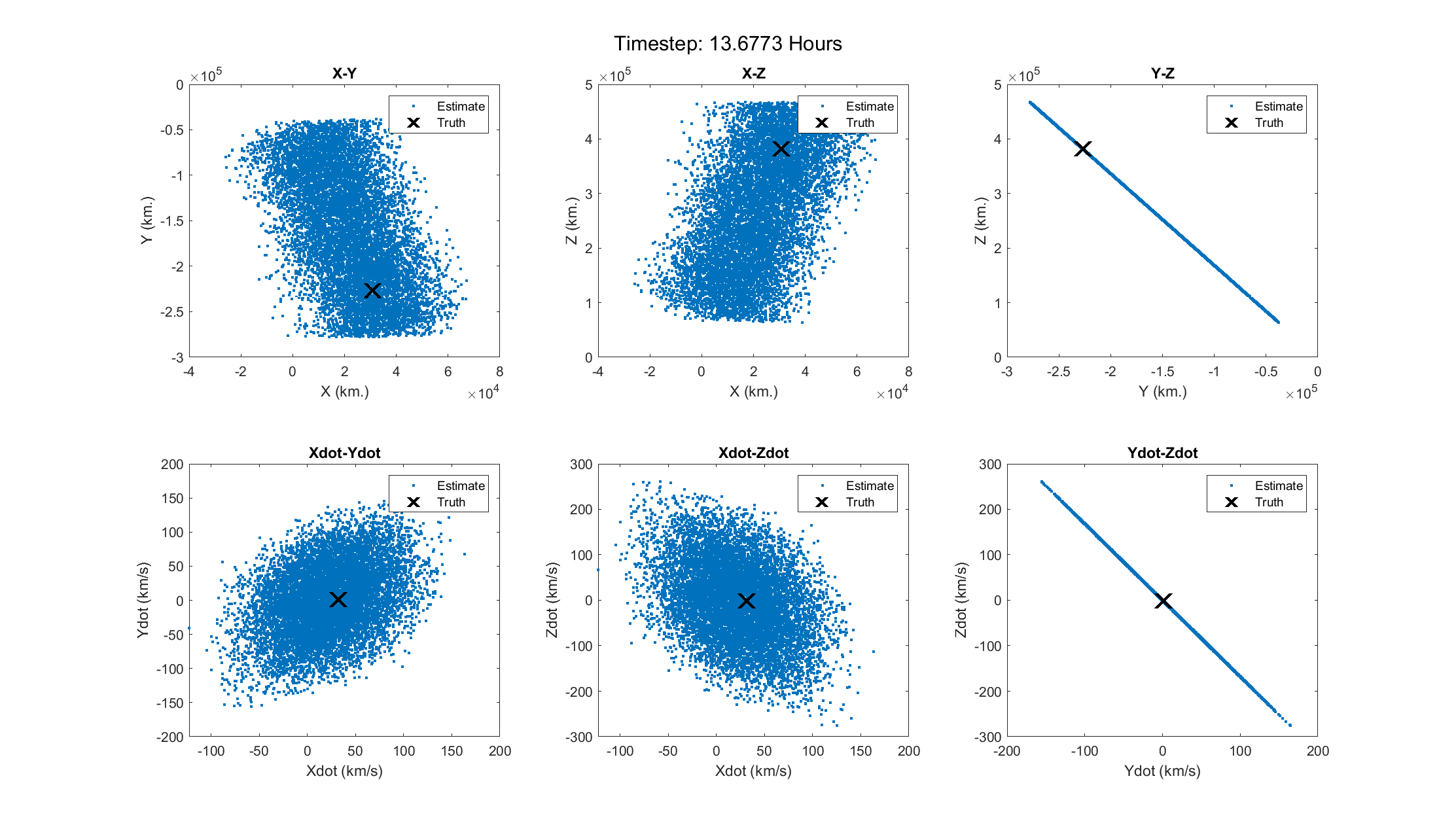}
	\caption{Particle representation of the initial state estimate of a target which passes through the $L_2$ Lagrange point at a very low velocity}
	\label{fig:8iodExample2}
\end{figure}

For this example, our initial state estimate is shown in particle form in Figure \ref{fig:8iodExample2}.

\begin{figure}[thpb] 
      \centering
      \begin{subfigure}{\textwidth}
            \centering
            \includegraphics[width=\linewidth]{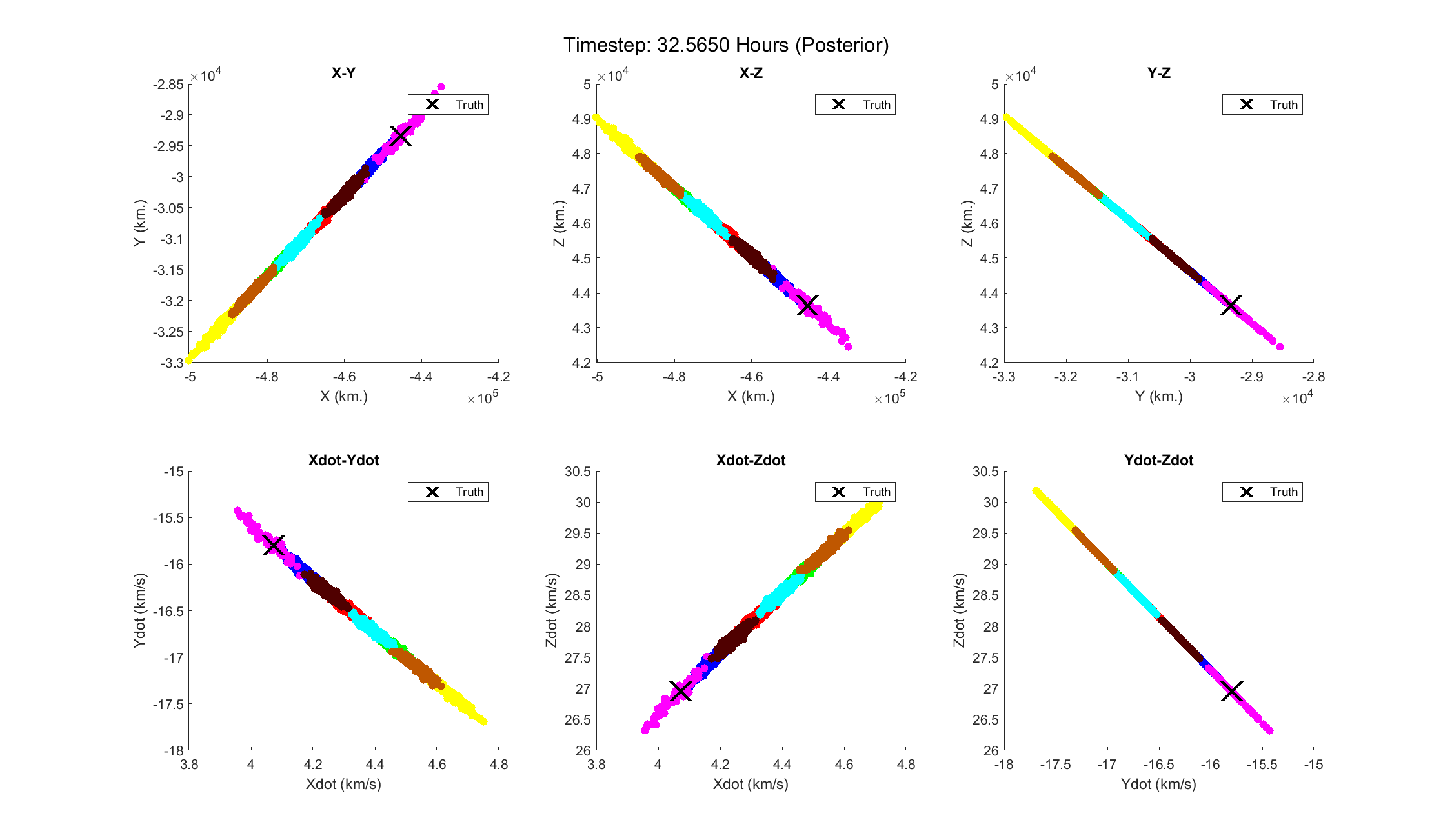}
            \caption{Target estimate just prior to shutoff} \label{fig:9a_endOf2nd}
      \end{subfigure}
      \vspace{0.5cm} % optional vertical spacing
      \begin{subfigure}{\textwidth}
            \centering
            \includegraphics[width=\linewidth]{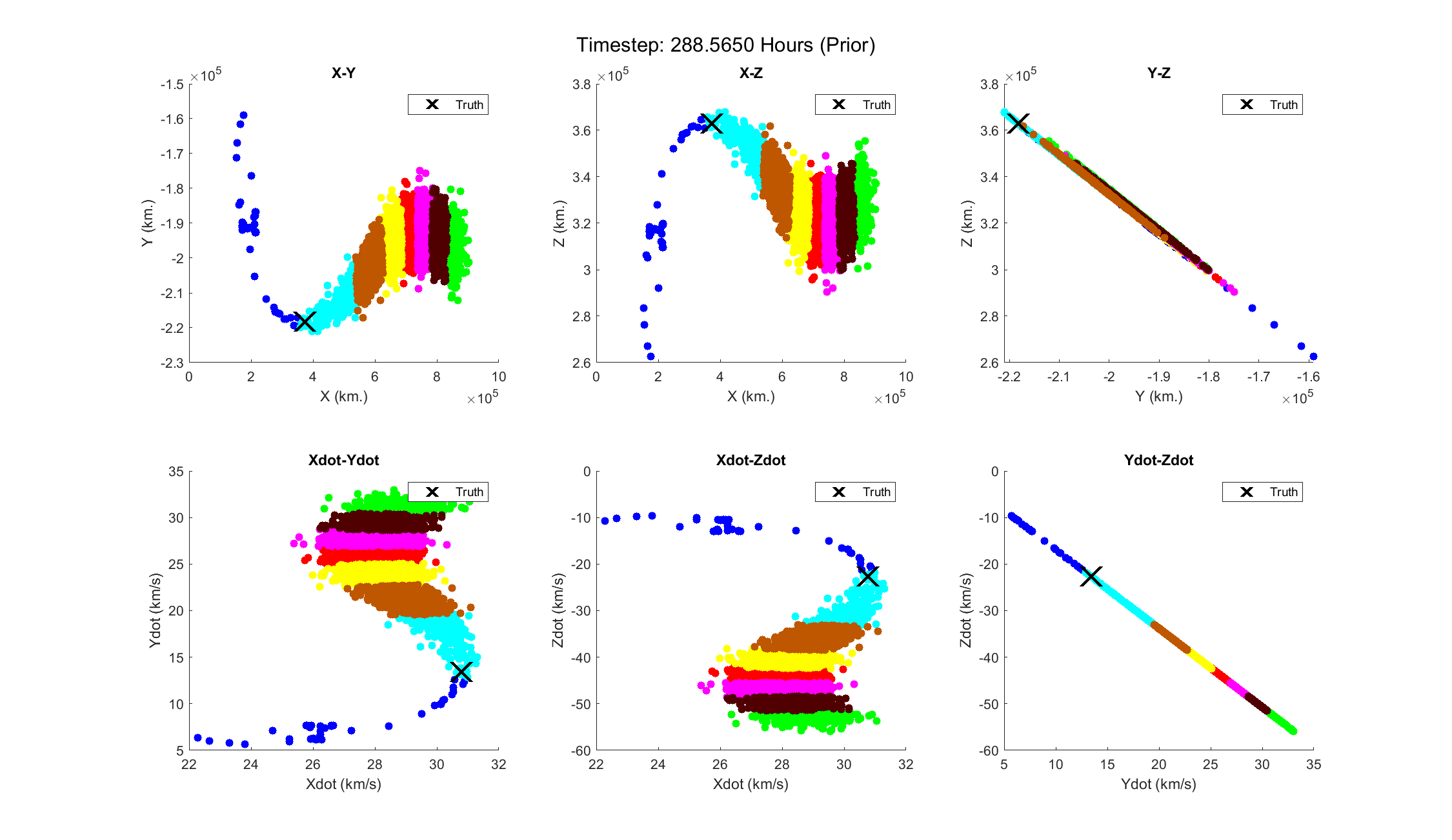}
            \caption{Target estimate once obtaining an observation is possible again}\label{fig:9b_endOfShutoff}
      \end{subfigure}

      \caption{Particle-based estimates of a target passing through the $L_2$ Lagrange point at a very small speed. The target estimate in (a) will warp chaotically, and the pencil-like PDFs will change shape into the tadpole-like PDFs in (b), which must be modeled as GMMs in order to successfully retain custody of the target.}
    \label{fig:9sampleL2results}
\end{figure}

Just as we accomplished for the first example, we utilize the PGM Filter to reduce the uncertainty of the initial state estimate. We receive angular measurements roughly every 40 minutes so long as our target is within our sensor FOV. Roughly 20 hours of PGM Filter iterations later, we induce a sensor shutoff that lasts approximately 10 days, enough for the Gaussianity of the state estimate to considerably break down. The PDFs of the target estimates prior to and post-shutoff (in clustered form) are given by Figure \ref{fig:9sampleL2results}.

\begin{figure}[h!]
	\centering\includegraphics[width=6in]{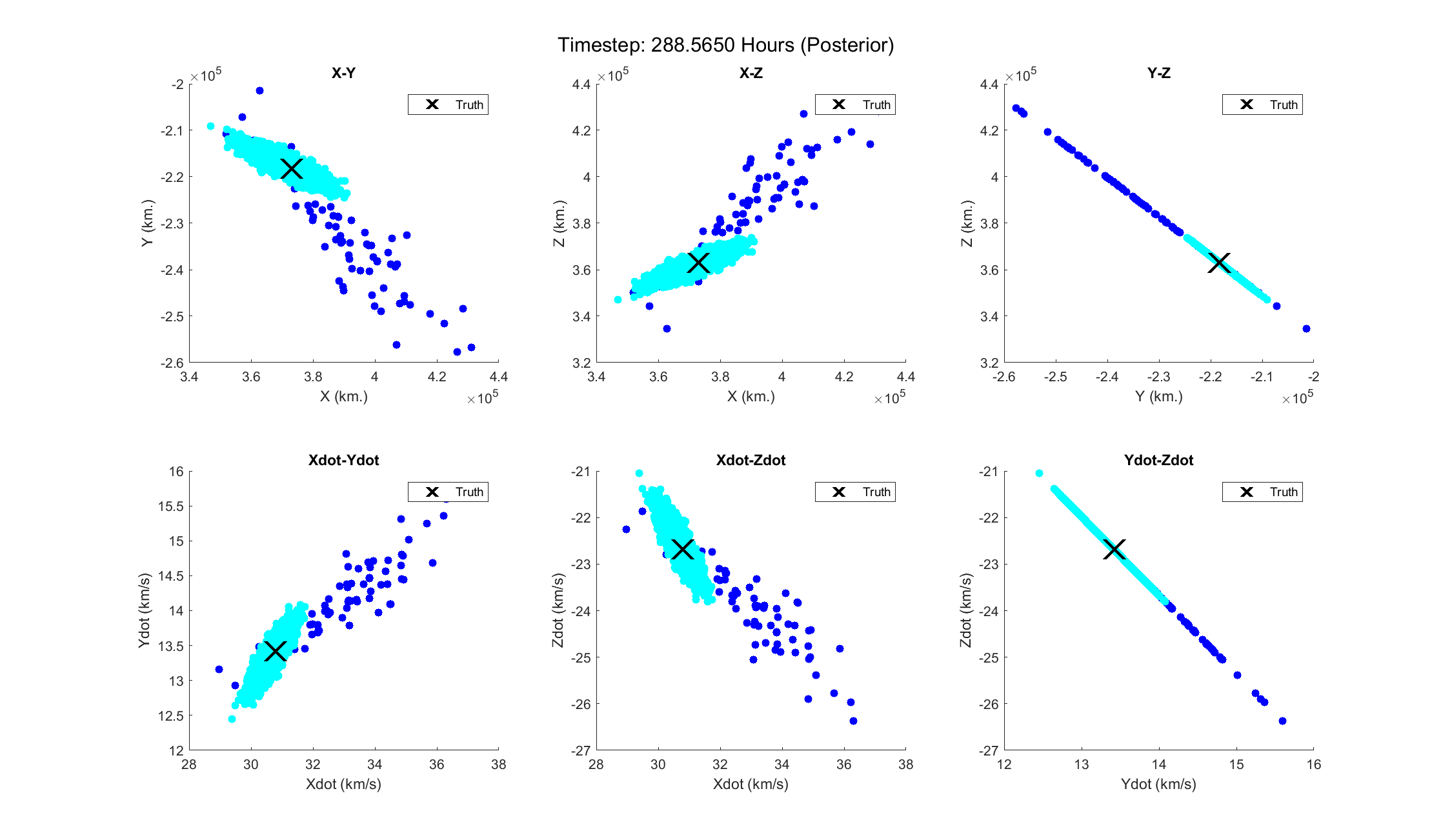}
	\caption{Posterior estimate of the target state after fusing a single observation after a ten-day sensor shutoff period which chaotically warped our Gaussian-like target state estimates in Figure \ref{fig:9b_endOfShutoff}.}
	\label{fig:10posteriorShutoff}
\end{figure}

We utilize the update step of the PGM Filter to fuse our first observation to eliminate most of the clusters in Figure \ref{fig:9b_endOfShutoff} and obtain a new, more precise state estimate in Figure \ref{fig:10posteriorShutoff}. Through this example, we demonstrate the ability of our IOD-OD framework to utilize minimal range information and a set of angles-only measurements to form an initial state estimate which, while very large, can be reduced by the PGM Filter, even in the cases of long sensor shutoffs.

\subsection{Example 3: Target Within the 3:1 NRHO Orbit}\label{subsec:ex3}

To further highlight the utility of our IOD-OD framework for chaotic cislunar orbits, we introduce a target in the 3:1 near-rectilinear Halo orbit around the $L_2$ Lagrange point. This orbit was chosen because of its growing interest within the cislunar astrodynamics community. Particularly, this orbit is said to be linearly stable in the CR3BP reference frame but "transition-challenging" when attempting to convert into the Higher-Fidelity Ephemeris Model (HFEM).\cite{park2024,davis2017} To demonstrate the robustness of IOD-OD framework to extreme cases, we propagate this target and its estimate through a nearly 150-day sensor shutoff period during which no observations are taken. In practice, in the event of such a long sensor shutoff, it would normally behoove us to reuse our initial orbit determination method over the first couple of observations post-shutoff to fit an initial state estimate. But even in this extreme scenario, we demonstrate that it is, in fact, possible to retain custody of our target after such a long shutoff without the need to perform initial orbit determination again.

\begin{figure}[h!]
	\centering\includegraphics[width=6in]{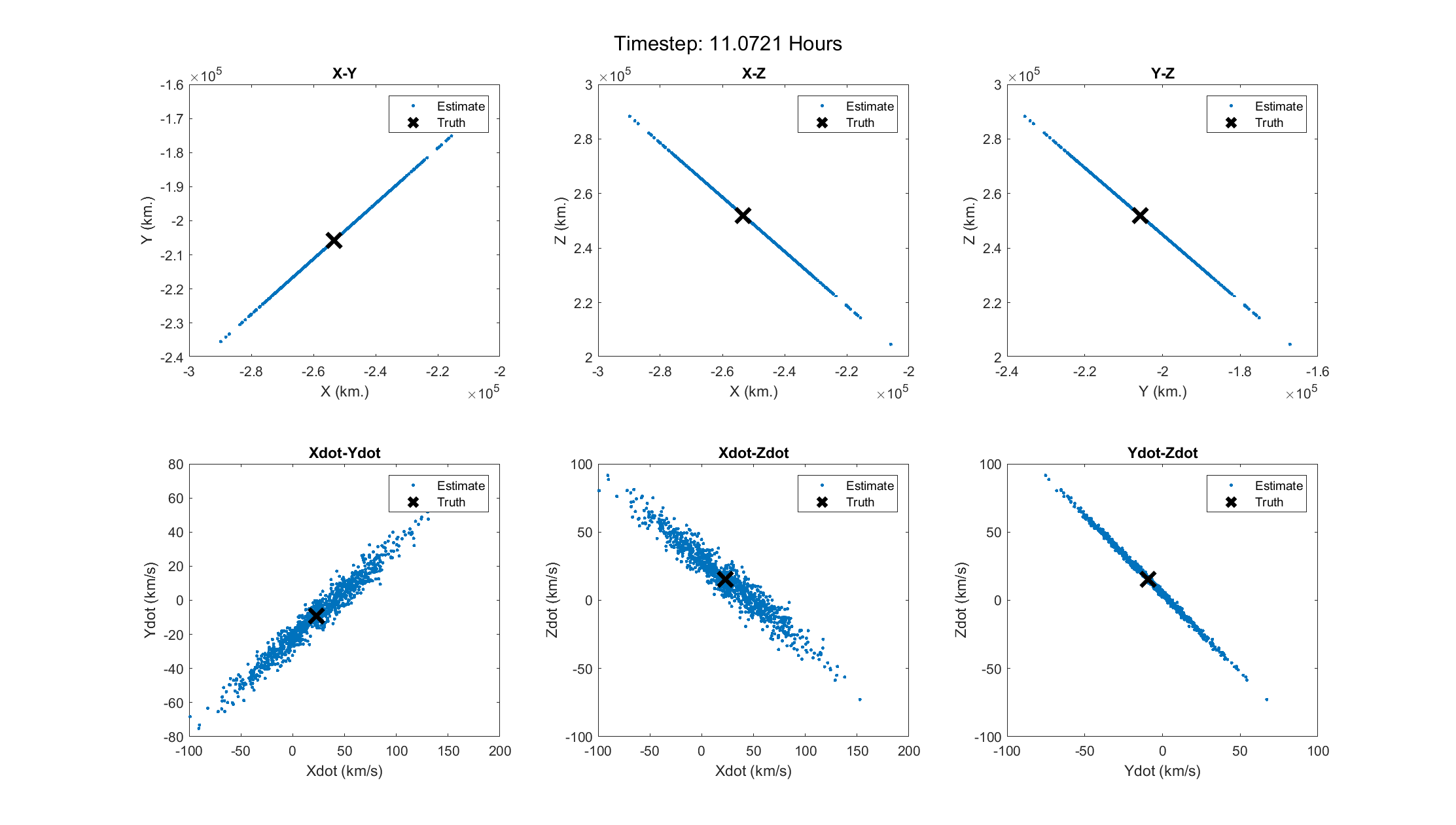}
	\caption{Particle representation of the initial state estimate of a target along the 3:1 NHRO. Unlike the previous examples, range is considered a measurement for IOD purposes, but the difficulty with obtaining a range estimate is modeled using a high range measurement noise, resulting in long, pencil-like PDFs.}
	\label{fig:11iodExample3}
\end{figure}

In Section \ref{subsec:IOD}, we list two different ways of obtaining range information with minimal assumptions. For the first two examples, we generated range values for each of our initial position estimates by drawing values from a large uniform PDF. In this example, however, we simply accept a noisy range "measurement", whose noise covariance matrix takes on a value of 0.05 for $\alpha$ from Eq. \ref{eq10:iod_R}. In other words, we assume that the range measurement of objects close to the Moon (roughly 400000 km. away from the Earth) will have a 20000 km. standard deviation to better reflect the difficulty of obtaining ground-based range estimates in cislunar space. Range measurements are not utilized after an initial state estimate has been generated. The resulting initial state estimate is given by Figure \ref{fig:11iodExample3}.

\begin{figure}[thpb] 
      \centering
      \begin{subfigure}{\textwidth}
            \centering
            \includegraphics[width=\linewidth]{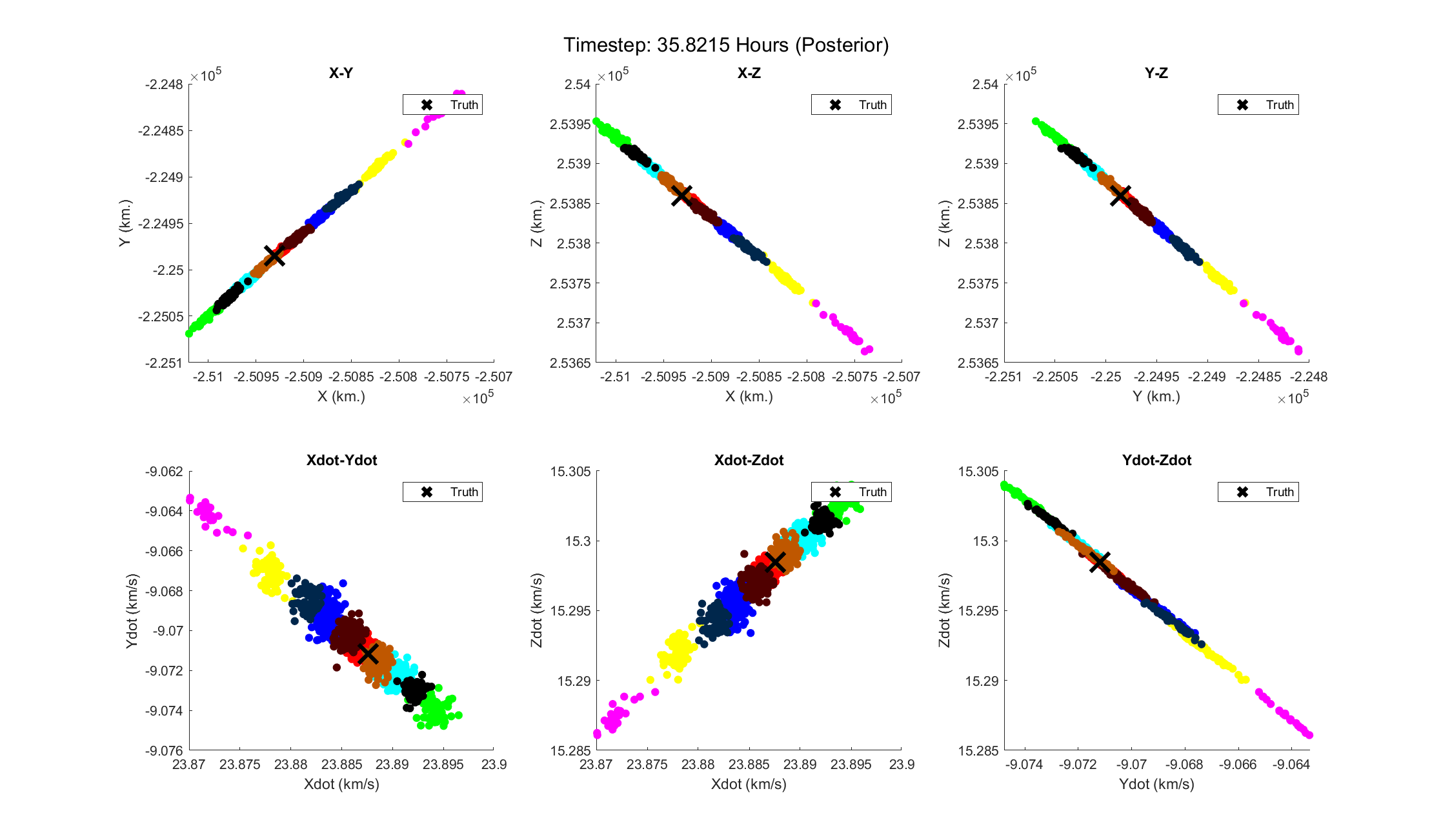}
            \caption{Target estimate just prior to shutoff} \label{fig:12a_endOf2nd}
      \end{subfigure}
      \vspace{0.5cm} % optional vertical spacing
      \begin{subfigure}{\textwidth}
            \centering
            \includegraphics[width=\linewidth]{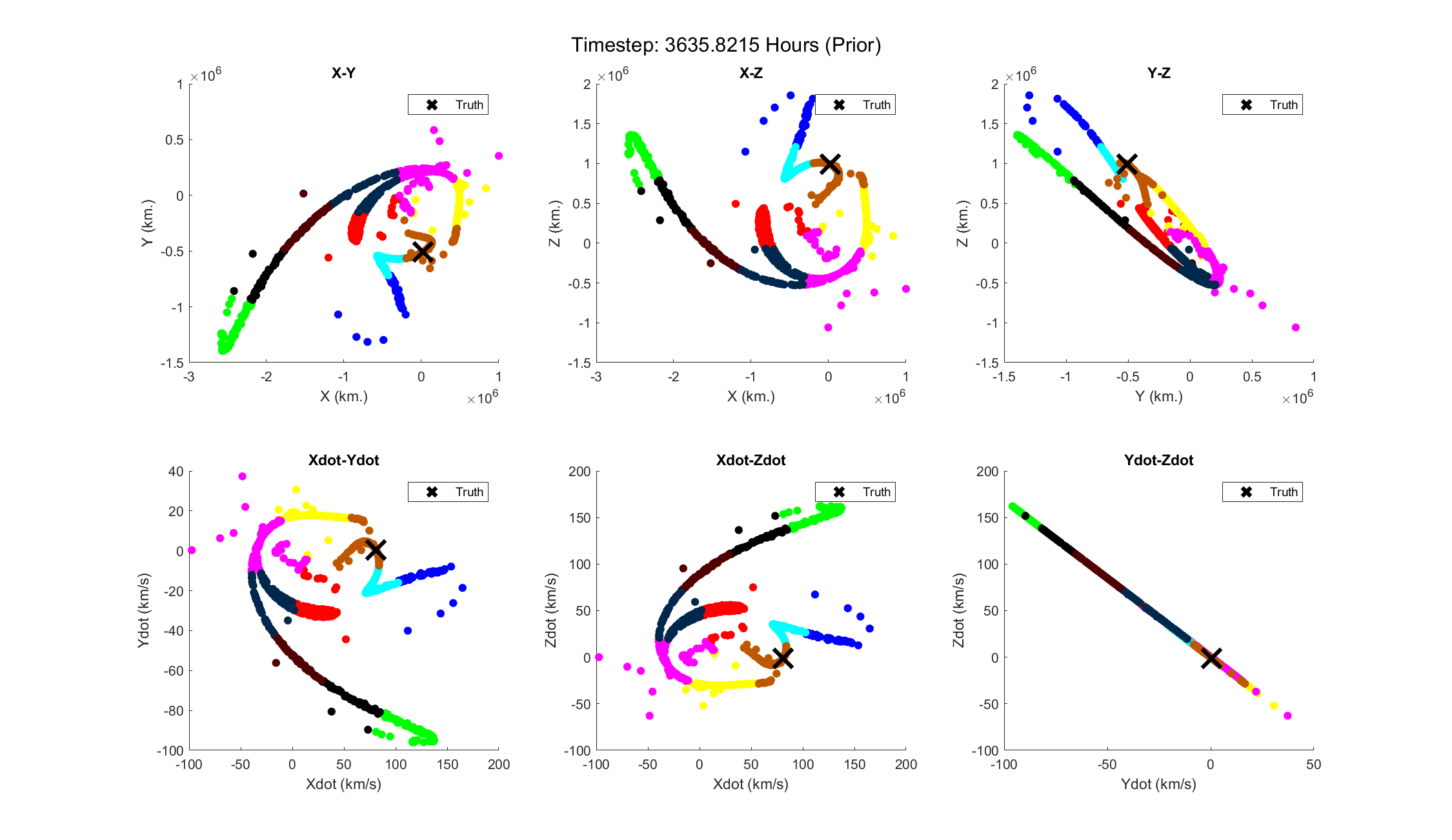}
            \caption{Target estimate once obtaining an observation is possible again. Sufficient warping and bifurcations are visible.}\label{fig:12b_endOfShutoff}
      \end{subfigure}

      \caption{Particle-based target estimates of a target within the 3:1 NRHO. The pre-shutoff target estimate in (a) warps and bifurcates chaotically, resulting in the PDF in (b). We utilize up to 10 clusters to accurately model this new state estimate.}
    \label{fig:12sample3-1results}
\end{figure}

Just as we had done in Section \ref{subsec:ex1}, we task our sensor to take observations roughly every 40 minutes for about 24 hours after IOD. Subsequently, we induce a 150-day sensor shutoff (15x larger than the second example) to allow our initial state estimate to warp and bifurcate even more. The pre- and post-shutoff state estimates are given by Figure \ref{fig:12sample3-1results}.

\begin{figure}[thpb] 
      \centering
      \begin{subfigure}{\textwidth}
            \centering
            \includegraphics[width=\linewidth]{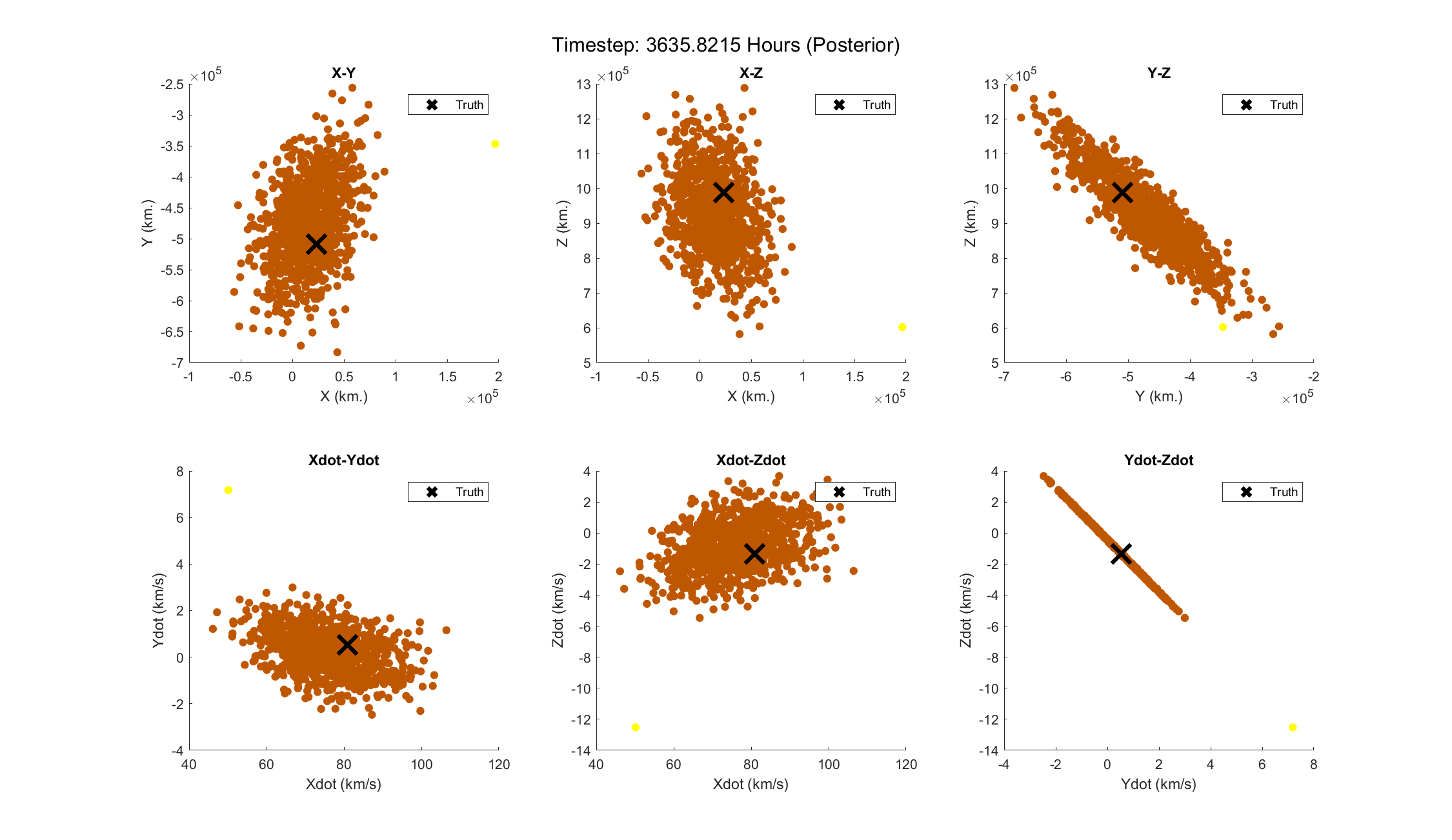}
            \caption{Target estimate once updated with an observation after a 150-day sensor shutoff} \label{fig:13a_ex3_postObsUpdate}
      \end{subfigure}
      \vspace{0.5cm} % optional vertical spacing
      \begin{subfigure}{\textwidth}
            \centering
            \includegraphics[width=\linewidth]{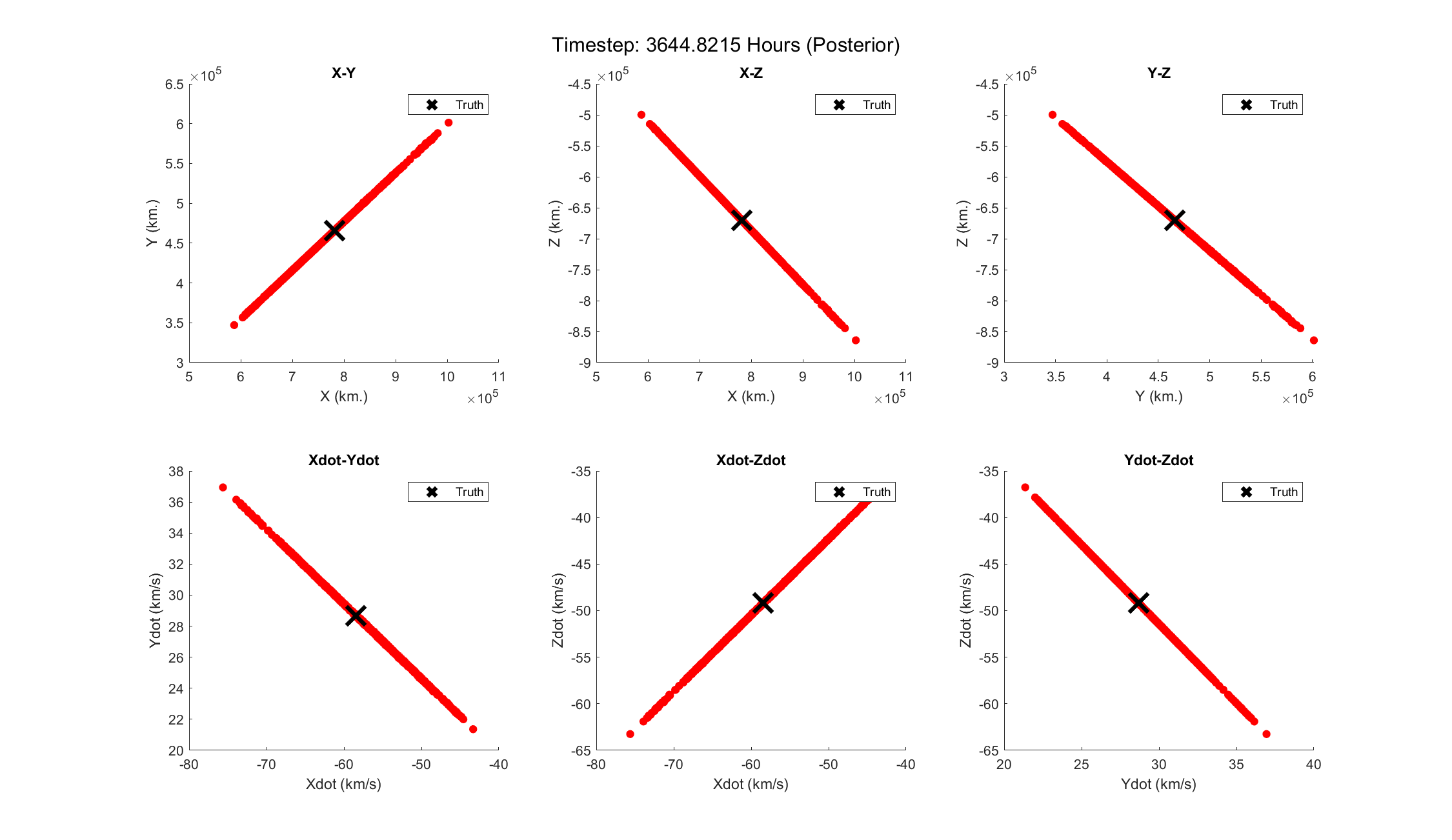}
            \caption{Target estimate two observation updates after the estimate in (a)}\label{fig:13b_3obsLater}
      \end{subfigure}

      \caption{Post-shutoff estimates of a target within the 3:1 NRHO after (a) a single observation update, and (b) after three observation updates. In spite of the low position precision in (a), the use of two more observations to obtain the estimate in (b) negates the need to perform IOD again.}
    \label{fig:13obsUpdate_results}
\end{figure}

Simply utilizing the PGM update step to Figure \ref{fig:12b_endOfShutoff} results in the posterior state estimate shown in Figure \ref{fig:13obsUpdate_results}. 

A quick inspection of Figure \ref{fig:13a_ex3_postObsUpdate} will show that the size of this posterior estimate is considerably larger than the initial state estimate in the position space, but the size of this posterior estimate is remarkably lower than the initial state estimate in the velocity space. Since our position estimate is not as precise as an IOD estimate at this point, we may think to utilize our IOD method to generate an initial position estimate while retaining our relatively precise estimate. However, this will require kinematically fitting polynomials through several observations. Instead, by updating our estimate with just two more observations, we reduce the precision of our position estimate to a similar precision as our initial state estimate (see Figure \ref{fig:13b_3obsLater}), rendering the need to perform IOD again unnecessary. Therefore, even in the event of abnormally long sensor shutoffs, the benefit of continuing to utilize the PGM Filter outweighs the need to perform IOD.

\subsection{Comparing Filter Performance for Cislunar Trajectories}\label{subsec:filterComp}

In Section \ref{subsec:ex3}, we demonstrated that perturbing a series of range, azimuth, and elevation observations with zero-mean, Gaussian measurement noise statistics and kinematically fitting polynomials through these measurement sets generally yields a smaller, continuous, and more compactly Gaussian initial state estimate. For this reason, it is practical to compare the performance of our PGM Filter relative to other, more well-known filters such as the Unscented Kalman Filter (UKF) and the Ensemble Kalman Filter (EnKF). Due to the orbit's importance, we compare filter performances with the 9:2 NRHO orbit and measurement schedule in Section \ref{subsec:ex1}, starting with the same initial state estimate given by Figure \ref{fig:5iodExample1} for all three filters, as well as the same observation schedule. We use the entropy metric defined in Eq. \ref{eq5:entropy}, noting that $M(k) = 1$ for all $k$ for the UKF and EnKF. 

\begin{figure}[h!]
	\centering\includegraphics[width=4in]{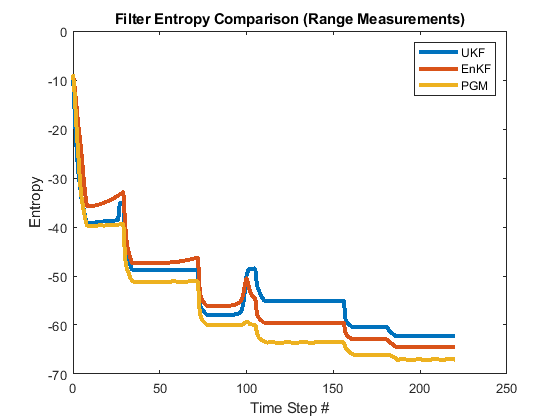}
	\caption{A comparison of filter performance (PGM, EnKF, and UKF) for the estimate of a target within the 9:2 NRHO for one full orbit starting with an IOD estimate resulting from weak range measurements.}
	\label{fig:14filterComp}
\end{figure}

Over each set of five measurements, the entropy is reduced. While entropy is theoretically expected to remain constant in between passes, slight jumps in the entropy exist in between passes due to the number of GMM components with which $p(\mathbf{x})$ from Eq. \ref{eq5:entropy} is approximated. Other entropy increases or jumps result from warping of the PDF. 

Unsurprisingly, the PGM Filter performs better than an EnKF. The PGM Filter utilizes the update step of an Ensemble Kalman Filter; indeed, the PGM Filter used throughout this article is a generalization of an EnKF to multiple Gaussian components. Although the EnKF update step is not as confident as an Extended Kalman Filter (EKF) update step, the PGM Filter entropy is lower post-update due to the fact that some or most of the GMM components will vanish or reduce in precision. As a result, the PGM Filter is consistently more confident than an EnKF.

The performance of the UKF with respect to the other filters is notable. For the first 2-3 passes, the performances of the PGM Filter and the UKF are almost indistinguishable. Since the unscented transform provides an approximate (and therefore, relatively less representative) representation for a target estimate than a PGM Filter (which uses a particle-based representation that becomes more accurate as the number of particles increases), the UKF update tends to become overconfident with time. The PGM Filter entropy tends to be lower due to the fact that multiple GMM components may vanish at the update step. Regardless, in the long term, the PGM Filter will more precisely track the target, while the UKF performance will eventually stabilize to a level almost indistinguishable from an EnKF.

\begin{figure}[thpb] 
      \centering
      \begin{subfigure}{\textwidth}
            \centering
            \includegraphics[width=\linewidth]{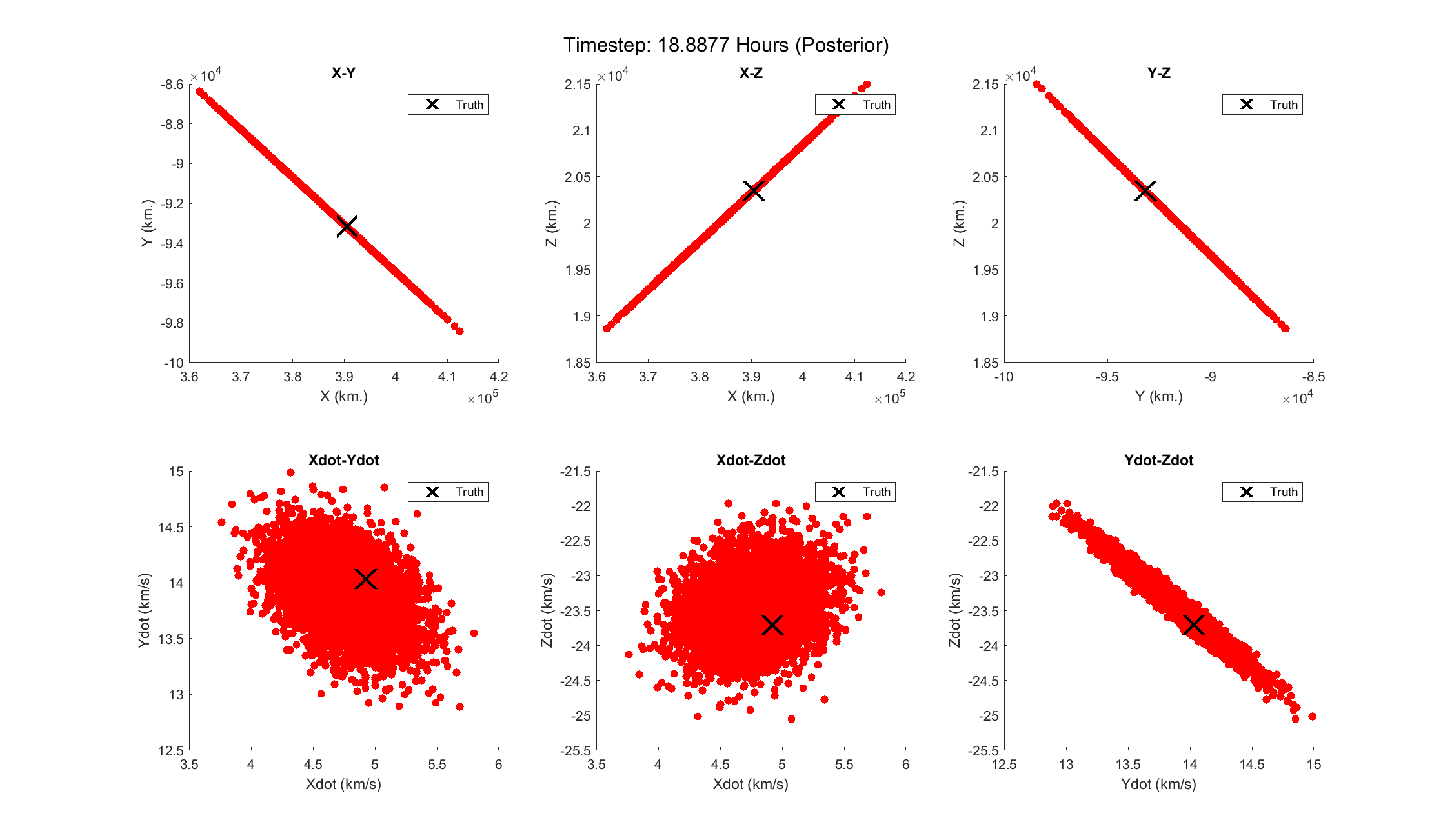}
            \caption{EnKF-based target estimate at the end of one pass} \label{fig:15a_enkf}
      \end{subfigure}
      \vspace{0.5cm} % optional vertical spacing
      \begin{subfigure}{\textwidth}
            \centering
            \includegraphics[width=\linewidth]{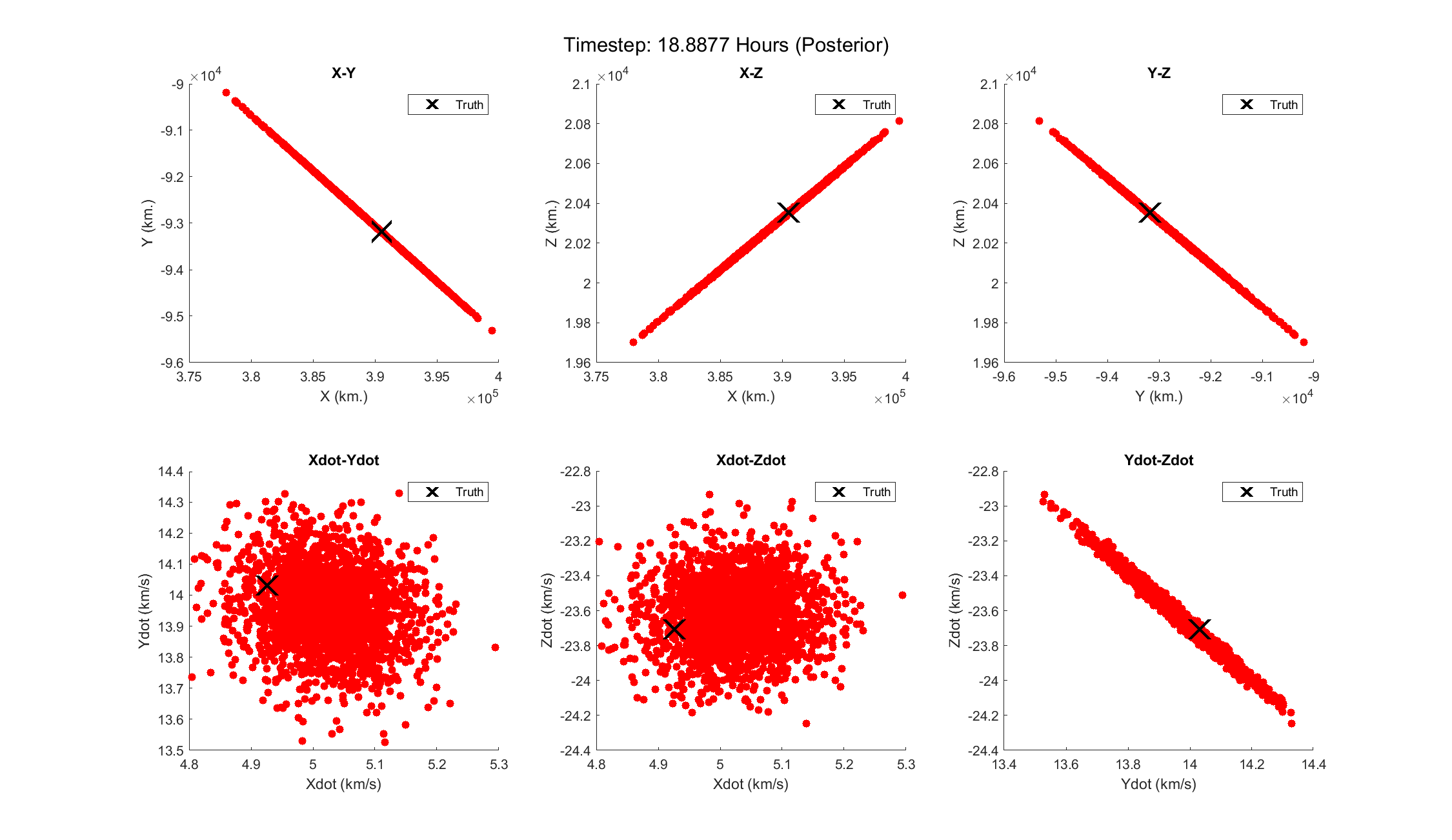}
            \caption{UKF-based target estimate at the end of one pass}\label{fig:15b_ukf}
      \end{subfigure}
      
      \caption{Particle-based representation of the confidence of a UKF relative to an EnKF for a target in the 9:2 NRHO after a single pass. The UKF-based target estimate is more overconfident than the EnKF-based target estimate due to how the unscented transform approximates posterior means and covariances. The state estimation error of the UKF compared to the EnKF is most prominent in the velocity space.}
    \label{fig:15enkf_v_ukf}
\end{figure}

Although Figure \ref{fig:14filterComp} suggests that a UKF performs better than an EnKF for almost the entirety of one full period of the 9:2 NRHO, the UKF's lower entropy results from the overconfidence of its update step.\cite{raihanukfpf2018} Among the UKF, EKF, and EnKF, the EnKF update step is the least confident. An advantage of this update step, however, is that the state estimation error relative to the $3\sigma$ intervals tends to be lower. By focusing on the target estimates at the end of the first pass of both the UKF and EnKF in Figure \ref{fig:15enkf_v_ukf}, we may observe that although the EnKF-based estimate of the target state is less precise, it is more accurate than the UKF-based estimate.

\begin{figure}[thpb] 
      \centering
      \begin{subfigure}{\textwidth}
            \centering
            \includegraphics[width=\linewidth]{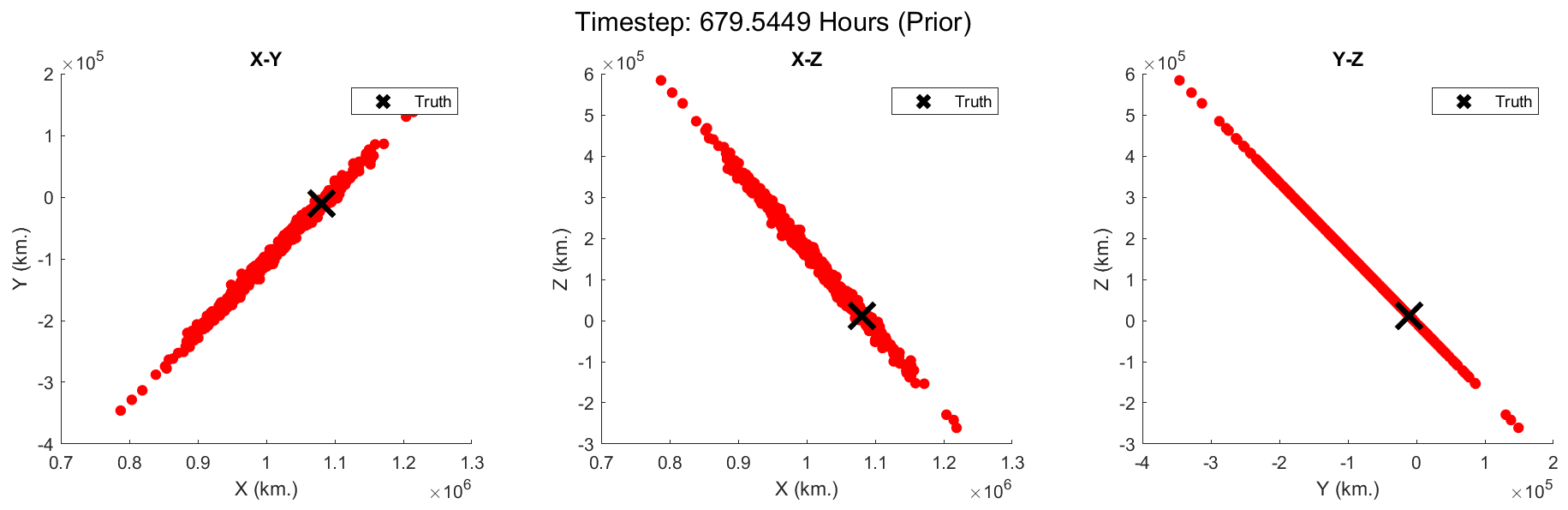}
            \caption{UKF-based target estimate at the end of a nearly 4-weeklong sensor shutoff} \label{fig:16a_ukf}
      \end{subfigure}
      \vspace{0.5cm} % optional vertical spacing
      \begin{subfigure}{\textwidth}
            \centering
            \includegraphics[width=\linewidth]{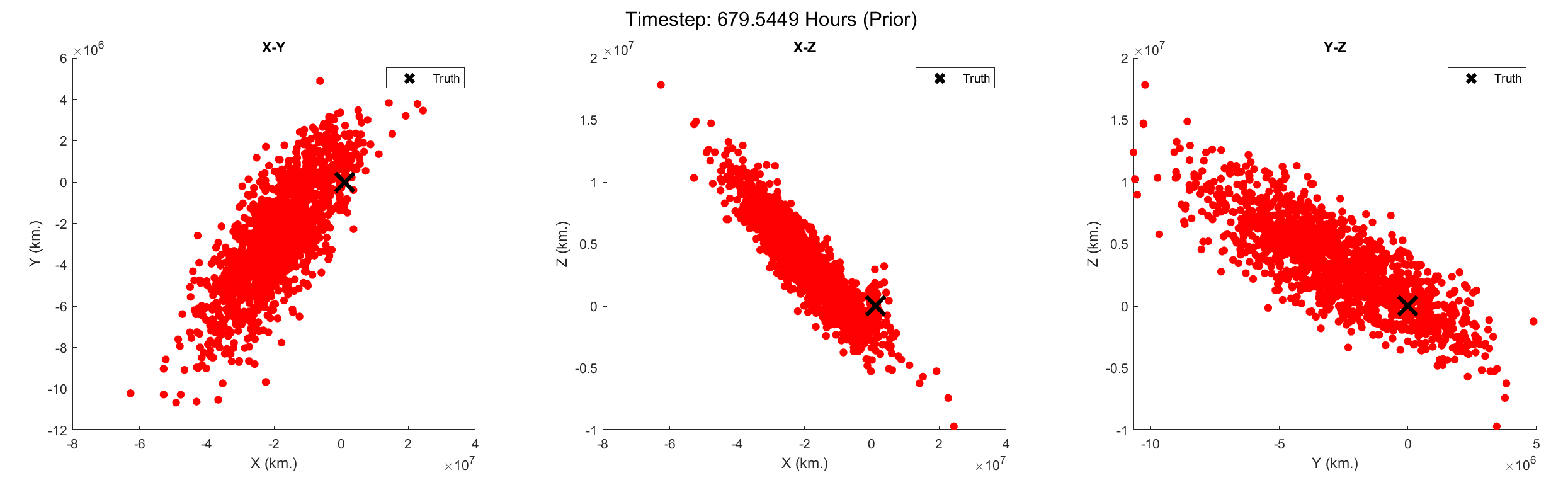}
            \caption{EnKF-based target estimate at the end of a nearly 4-week long sensor shutoff}\label{fig:16b_enkf}
      \end{subfigure}
      \vspace{0.5cm} % optional vertical spacing
      \begin{subfigure}{\textwidth}
            \centering
            \includegraphics[width=\linewidth]{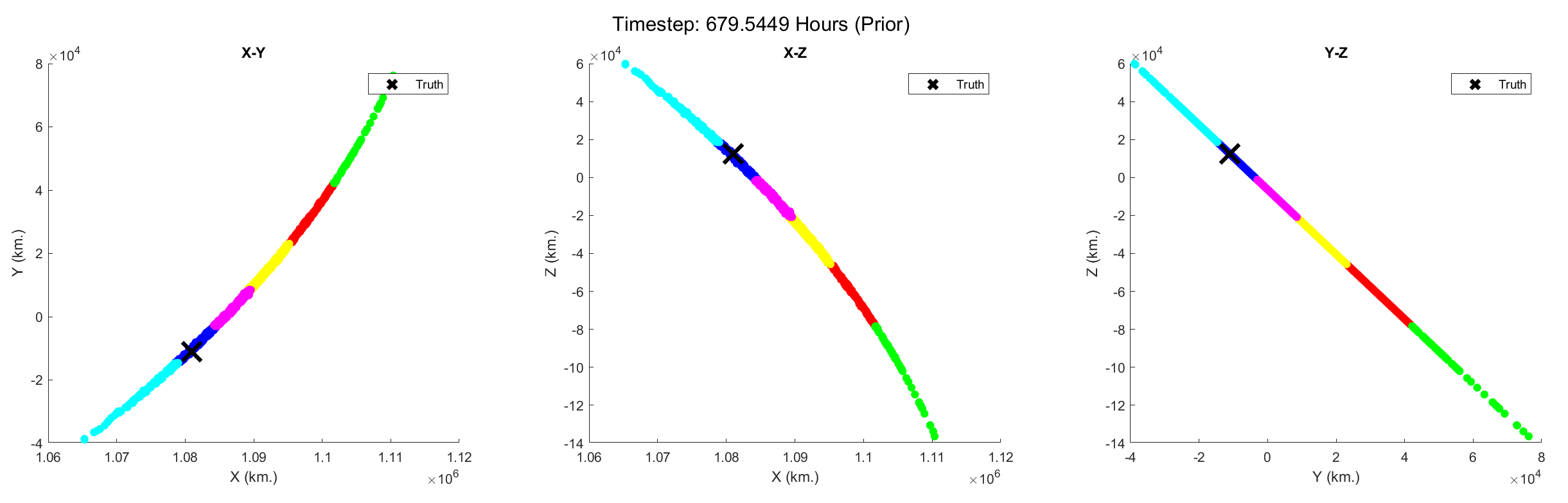}
            \caption{PGM-based target estimate at the end of a nearly 4-week long sensor shutoff}\label{fig:16c_pgm}
      \end{subfigure}
      
      \caption{Post-shutoff \textit{a priori} state estimates of a target passing through the $L_2$ Lagrange point at a low speed (see Section \ref{subsec:ex2}). Light to moderate warping is observed depending on the filter.}
    \label{fig:16preShutoff_filters}
\end{figure}

The above example demonstrates that although the PGM Filter outperforms the UKF and EnKF, the difference is not significant. A key reason for this is that the initial state estimate and subsequent \textit{a priori} estimates are able to be modeled as approximately Gaussian due to how we model range information and our measurement noise characteristics. On top of that, the propagation time steps between observations are not significant enough to observe a breakdown in the Gaussianity of the posterior target estimate at a previous time step. To that end, we shall revisit the case of moderate breakdown in Gaussianity observed in Section \ref{subsec:ex2}. Following an almost 4-week long shutoff, each of our three filtering techniques resulted in the target estimates shown in Figure \ref{fig:16preShutoff_filters}.

\begin{figure}[thpb] 
      \centering
      \begin{subfigure}{\textwidth}
            \centering
            \includegraphics[width=\linewidth]{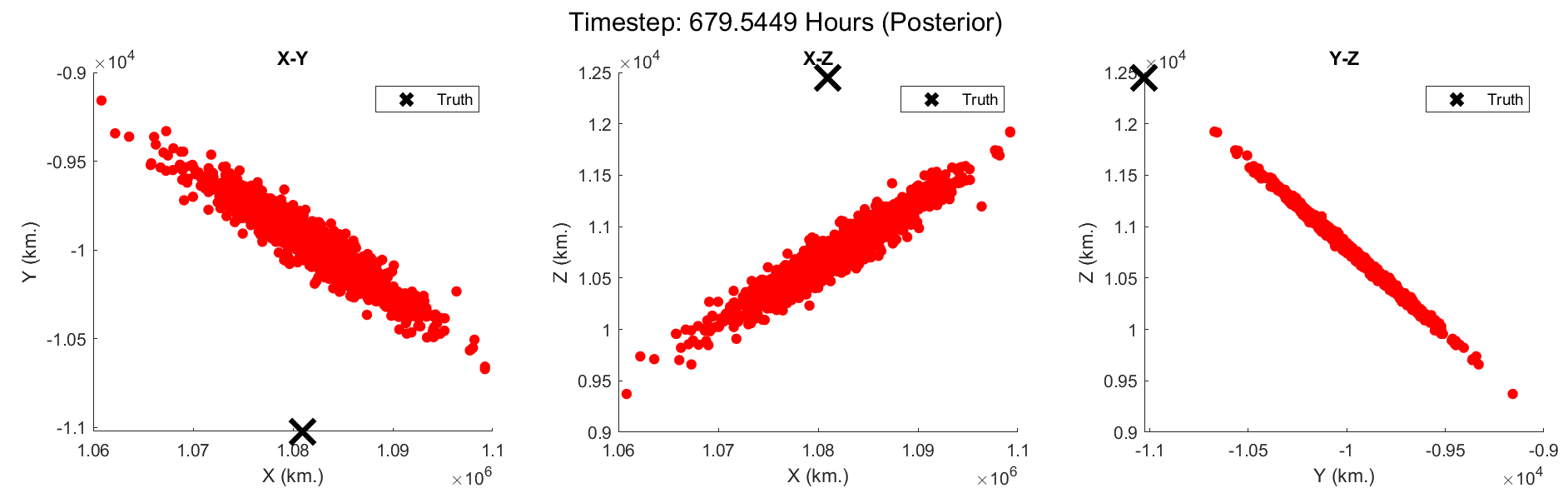}
            \caption{UKF-based target estimate after a single observation update from Figure \ref{fig:16a_ukf}} \label{fig:17a_ukfFailure}
      \end{subfigure}
      \vspace{0.5cm} % optional vertical spacing
      \begin{subfigure}{\textwidth}
            \centering
            \includegraphics[width=\linewidth]{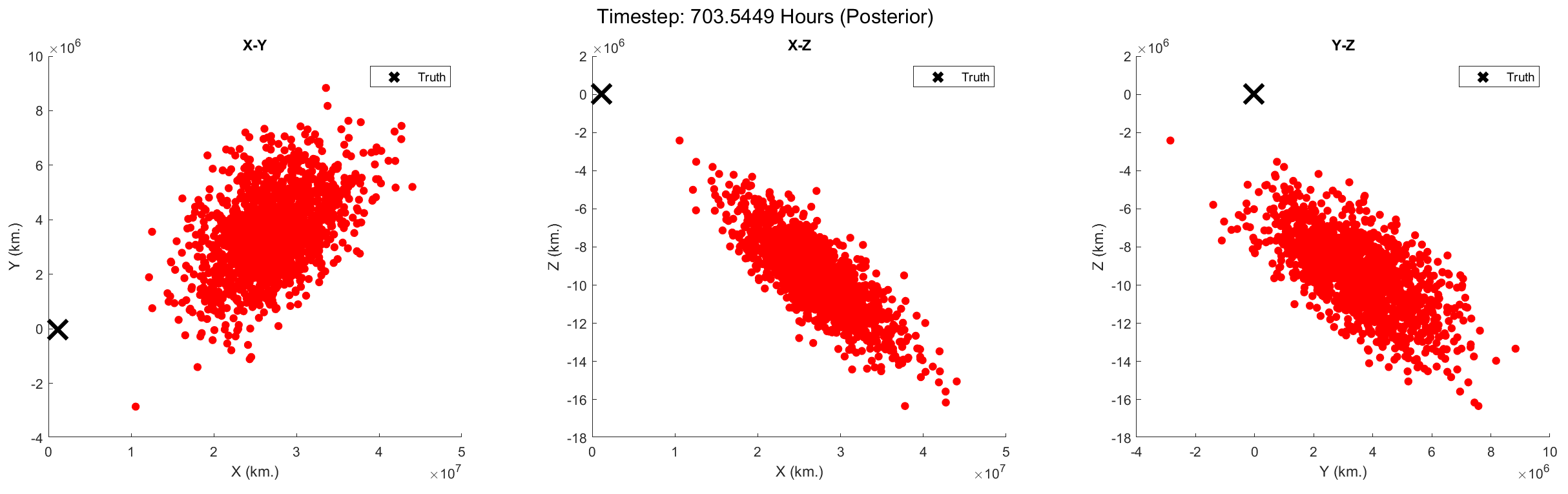}
            \caption{EnKF-based target estimate after a few observation updates from Figure \ref{fig:16b_enkf}}\label{fig:17b_enkfFailure}
      \end{subfigure}
      \vspace{0.5cm} % optional vertical spacing
      \begin{subfigure}{\textwidth}
            \centering
            \includegraphics[width=\linewidth]{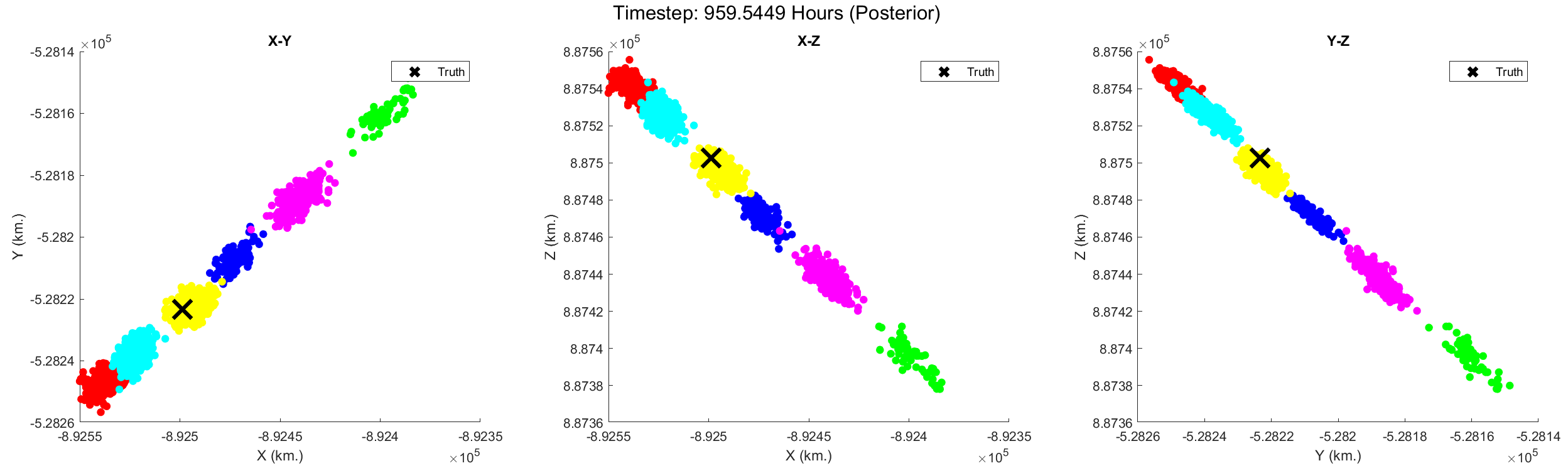}
            \caption{PGM-based target estimate consistency nearly two weeks after the long shutoff period}\label{fig:17c_pgmSuccess}
      \end{subfigure}
      
      \caption{Points at which the UKF and EnKF filter estimates become inconsistent with the target states. Only the PGM Filter is capable of maintaining custody of the target in the long term post-shutoff.}
    \label{fig:17_filterFails}
\end{figure}

The explosive growth of the EnKF-based estimate relative to the PGM-based and UKF-based estimate in Figure \ref{fig:16preShutoff_filters} may be attributed to the chaotic divergence of trajectories within close proximity of the $L_2$ Lagrange points and the lower confidence update steps of an EnKF prior to the shutoff. Although this does not immediately cause the EnKF-based estimate to become inconsistent after a single observation update, the unwieldiness and size of the resulting PDFs will eventually cause an inconsistent update after a few more filter iterations, as shown in Figure \ref{fig:17_filterFails}. On the other hand, the overconfidence of the UKF update step will cause the filter's estimate of the target state to immediately become inconsistent after this long shutoff. Only the PGM Filter is resilient to this chaotic warping for the long term.

Although the UKF and EnKF serve as appropriate filters for more stable, observation-frequent, and compact-estimate trajectories within cislunar space, the PGM Filter is much more robust at maintaining custody of targets in more chaotic trajectories. For long sensor shutoff periods, the PGM Filter will outshine both  UKF and EnKF filters. 

% \textcolor{red}{Are the UKF/ EnKF always viable for observation frequent scenarios? For instance, do they work for the "uniform in cislunar range" case?}

\subsection{Range Information Assumptions and Their Effect on IOD}\label{subsec:assumptionComp}

In this section so far, we have discussed and demonstrated the effects of two assumptions of range information and their effect on the initial state estimate of a target. The first assumption of range information involved assuming any range distributed as $\rho \sim \mathcal{U}[84328, 550000]$ km., roughly corresponding to the range boundaries of cislunar space. The second assumption involved taking range measurements with an assumption of high measurement noise (e.g. nominally 5\% of the true range). Although the resulting initial state estimates (see Figures \ref{fig:5iodExample1}, \ref{fig:8iodExample2}, and \ref{fig:11iodExample3}) are consistent with the target truth, they are massive, especially in the velocity space. Most RSOs and targets of interest in cislunar space are expected to stay in cislunar space (especially if they are in any type of orbit around the Earth-Moon system). Therefore, we are able to make the reasonable assumption that the target is not actively attempting to escape the solar system, and we can therefore filter out any state estimates whose velocity magnitude is greater than 42 km/s. Shifting our attention again to the 9:2 NRHO trajectory from \ref{subsec:ex1}, we demonstrate the effect of these three sets of assumptions on the initial state estimate, namely: 1) Noisy range measurements, 2) range bounding only, and 3) range and velocity bounding. 

\begin{figure}[thpb] 
      \centering
      \begin{subfigure}{\textwidth}
            \centering
            \includegraphics[width=0.66\linewidth]{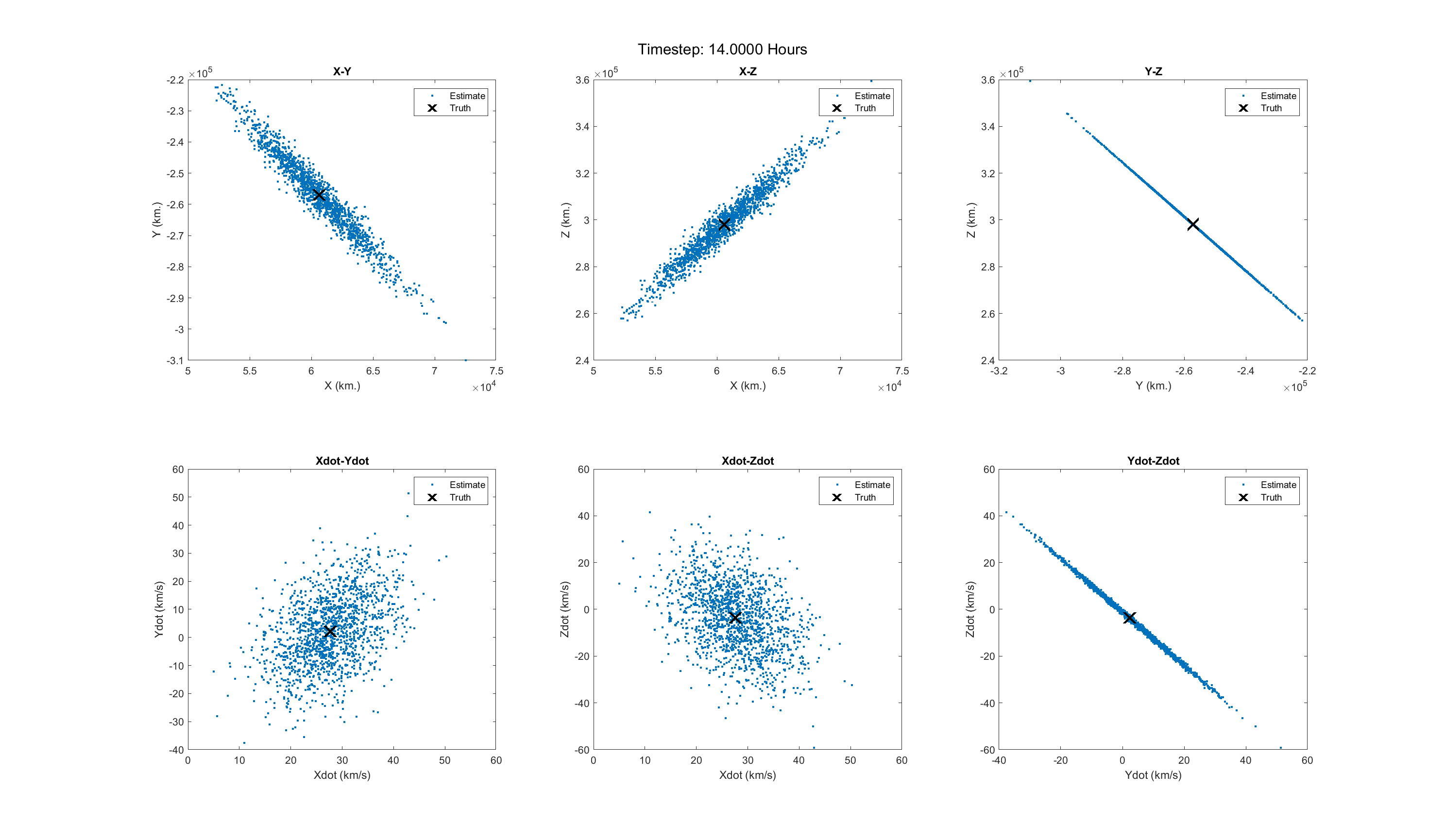}
            \caption{IOD resulting from drawing several range measurements with a 5\% standard deviation and kinematically fitting with several observations.} \label{fig:19a_rangeMeas}
      \end{subfigure}
      \vspace{0.5cm} % optional vertical spacing
      \begin{subfigure}{\textwidth}
            \centering
            \includegraphics[width=0.66\linewidth]{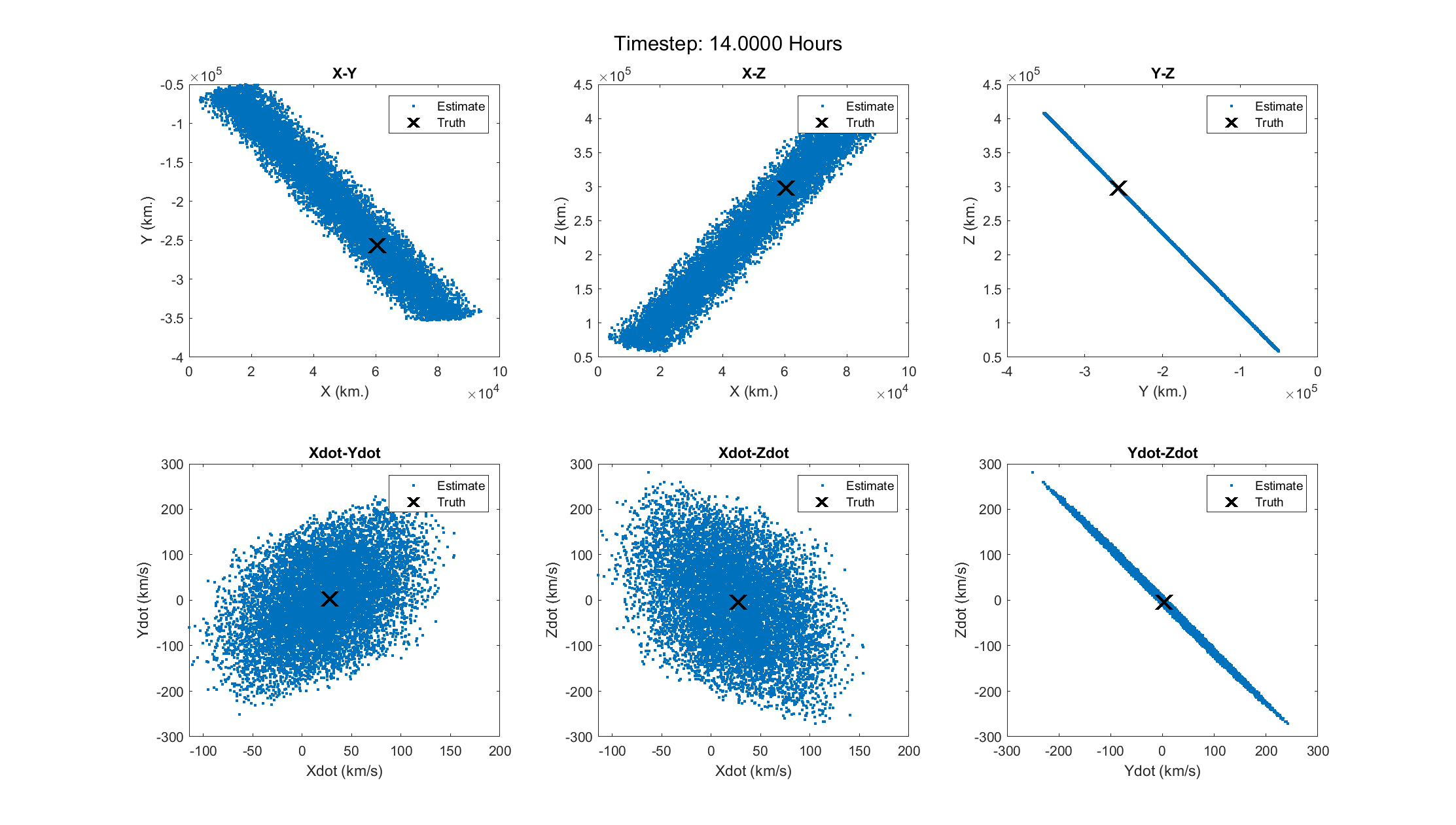}
            \caption{IOD resulting from drawing range from the PDF $\rho \sim \mathcal{U}[84328, 550000]$ km. and kinematically fitting with several observations.}\label{fig:19b_rangeOnly}
      \end{subfigure}
      \vspace{0.5cm} % optional vertical spacing
      \begin{subfigure}{\textwidth}
            \centering
            \includegraphics[width=0.66\linewidth]{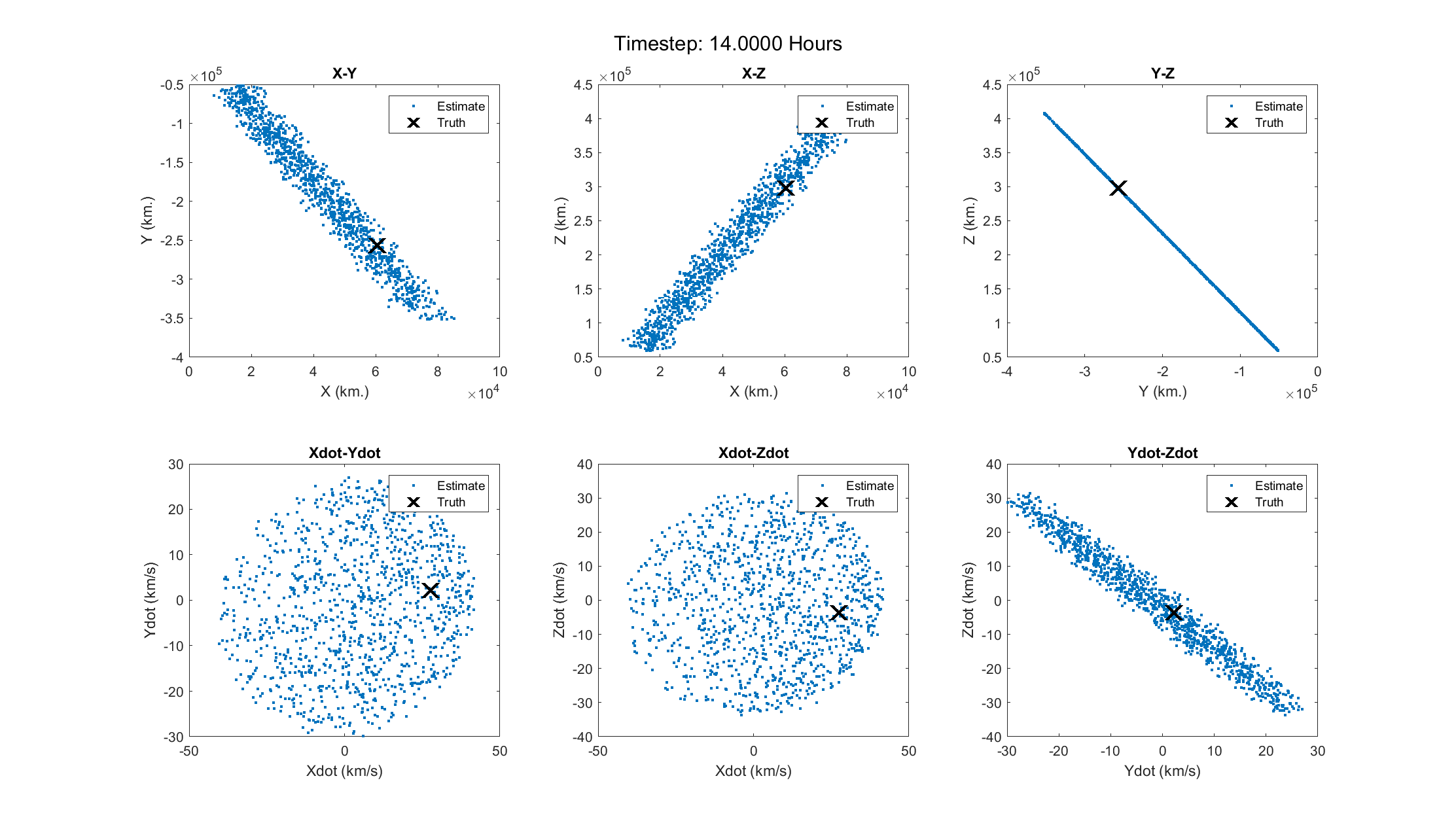}
            \caption{IOD resulting from drawing range from the PDF $\rho \sim \mathcal{U}[84328, 550000]$ km., kinematically fitting with several observations, and filtering out particles below a speed of 42 km/s.}\label{fig:19c_rangeVel}
      \end{subfigure}
      
      \caption{Initial state estimates resulting from three sets of assumptions: 1) Noisy range measurements, 2) range bounding only, and 3) range and velocity bounding.}
    \label{fig:19_iodAssumptions}
\end{figure}

\begin{figure}[h!]
	\centering\includegraphics[width=6in]{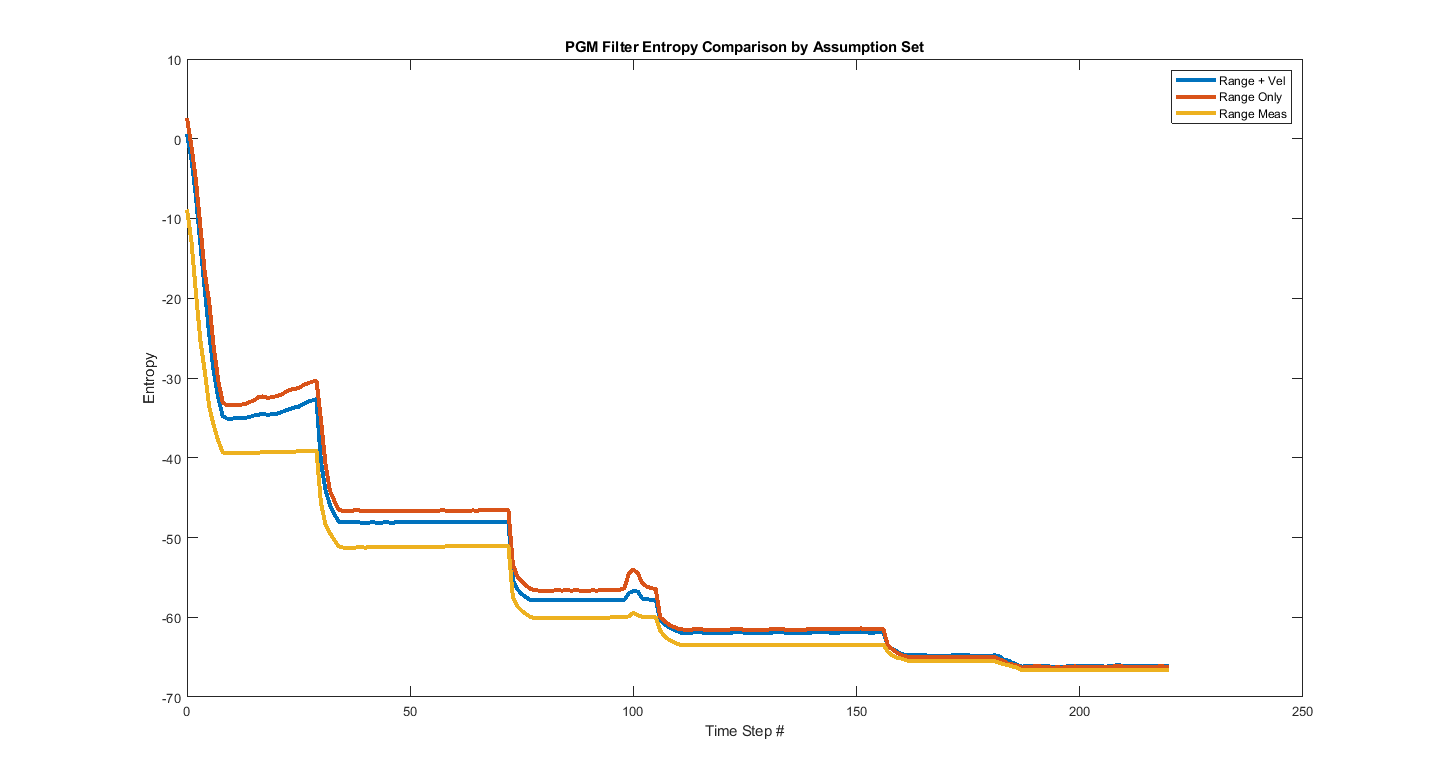}
	\caption{A comparison how well we can estimate the state of a target within the 9:2 NRHO over one full orbit based on three different sets of assumptions about the range and/or velocity: 1) Noisy range measurements ("Range Meas"), 2) range bounding only ("Range Only"), and 3) range and velocity bounding ("Range + Vel").}
	\label{fig:18assumptionComp}
\end{figure}

Since we assume additive, zero-mean Gaussian measurement noise for the assumption set corresponding to Figure \ref{fig:19a_rangeMeas}, however large, the target truth is generally expected to lie close to or at the center of the initial state estimate. Although the range noise is high, it is not nearly large enough to encompass the cislunar domain, yielding us an IOD estimate that is much smaller than what is obtained with our range bounding and range and velocity bounding assumptions. The assumption set involving range and velocity bounding builds upon the assumption set of range bounding only by adding a velocity constraint. Understandably, this constraint results in the filtering of many of the particles in the bottom three subplots of Figure \ref{fig:19b_rangeOnly} and yield a much smaller estimate, especially in the velocity space. Due to the size of the PDF from which we draw range estimates in Figures \ref{fig:19b_rangeOnly} and \ref{fig:19b_rangeOnly}, the initial velocity estimates, although consistent with the true velocity, become massive. With the velocity constraint, up to 85\% of the particles from Figure \ref{fig:19b_rangeOnly} can get filtered out, leaving a dramatically sparser, smaller set of particles such as those in Figure \ref{fig:19c_rangeVel}. Although techniques such as subset simulation may be used to more efficiently and iteratively reduce the state estimate resulting from range bounding, we compensate for the large loss of particles by performing IOD with nearly 5-10 times as many particles as we wish to use for OD. Once the initial ensemble of particles is propagated and updated with a measurement, the resampling step would sample the desired number of particles from the posterior GMM distribution. To observe the long-term performance of the PGM Filter upon the three IOD estimates in Figure \ref{fig:19_iodAssumptions}, we utilize the entropy metric from Eq. \ref{eq5:entropy} to obtain Figure \ref{fig:18assumptionComp}. Once again, we utilize the 9:2 NRHO orbit with a similar schedule of measurements detailed in Sections \ref{subsec:ex1} and \ref{subsec:filterComp}.

Although the three entropy curves start at different points along the $y$-axis, they converge to similar entropy values. In effect, the number of assumptions utilized for IOD will only affect the precision of the target estimate in the short term -- typically the first one or two passes. If the objective is to track a target in cislunar space for the long term, then a minimal number and type of range assumptions (i.e. those which correspond to Figure \ref{fig:19b_rangeOnly}) is sufficient. 

\subsection{Comparison of Filter Performances with Differing IOD Estimates}\label{subsec:compareAll}

In Section \ref{subsec:filterComp}, we compared the effects of two more common filtering frameworks with the PGM Filter starting with an initial state estimate corresponding to the assumption of range measurements ("Range Meas" from Figure \ref{fig:18assumptionComp}). In Section \ref{subsec:assumptionComp}, we identified three sets of assumptions that would result in differing IOD estimates, and explored how much the PGM Filter would assist in the short and long-term entropy evolution for each initial state estimate. In this section, we combine results from Sections \ref{subsec:filterComp} and \ref{subsec:assumptionComp} in an effort to further differentiate the performance of the PGM Filter with our kinematic fitting framework, focusing on the latter two assumption sets from Section \ref{subsec:assumptionComp}.

\begin{figure}[h!]
	\centering\includegraphics[width=0.5\linewidth]{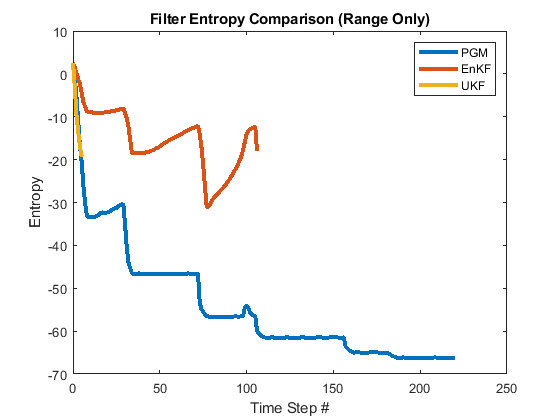}
	\caption{A comparison how well we can estimate the state of a target within the 9:2 NRHO over one full orbit based on an initial state estimate derived from the minimal range information assumption set of Section \ref{subsec:assumptionComp}. The PGM Filter is compared with the EnKF and UKF in the same manner as Figure \ref{fig:14filterComp}.}
	\label{fig:21assumptionComp_rangeOnly}
\end{figure}

First, we focus on the minimal range assumption, corresponding to the "Range Only" curve in Figure \ref{fig:18assumptionComp}. To briefly summarize, our kinematic framework incorporates range information by drawing from a large, uniformly-distributed PDF encompassing all valid ranges of the cislunar domain \textit{without} filtering by speed. The resulting initial state estimate is a large, often unwieldly and non-Gaussian PDF. Figure \ref{fig:21assumptionComp_rangeOnly} demonstrates what happens when we utilize a UKF, EnKF, and the PGM Filter in conjunction with the IOD estimate corresponding to the minimal range assumption set. 

\begin{figure}[h!]
	\centering\includegraphics[width=0.5\linewidth]{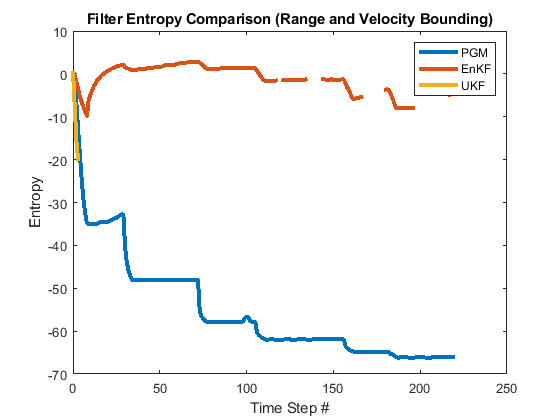}
	\caption{A comparison how well we can estimate the state of a target within the 9:2 NRHO over one full orbit based on an initial state estimate derived from the minimal range information with velocity bounding assumption set of Section \ref{subsec:assumptionComp}.}
	\label{fig:22assumptionComp_rangeVel}
\end{figure}

Compared to Figure \ref{fig:14filterComp}, there is a much starker difference in the performance of the PGM Filter with respect to either the UKF or the EnKF. In fact, in this measurement scheduling scenario, only the PGM Filter is capable of precisely and accurately estimating the target state over one full orbit. Due to the size and unwieldiness of the initial state estimate, a UKF slowly becomes overconfident in its subsequent state estimates. As a result, the filter fails to maintain custody of the target within the first pass itself. Due to the size and shape of the initial state estimate, an EnKF is much slower at reducing  the initial state estimate during the RSO's first pass. Due to the estimate size and chaotic warping effects between the third and fourth passes, the EnKF entropy estimate sees a huge spike over several time steps. Once a measurement schedule is obtained again, the size and shape of the estimate become so unwieldy that the EnKF soon fails to retain custody of the target. However, the PGM Filter is robust to these effects, and outperforms the UKF and EnKF for this scenario as well. A similar behavior is observed with the addition of a velocity constraint (corresponding to the "Range + Vel" curve of Figure \ref{fig:18assumptionComp}). The corresponding entropy evolution is shown in Figure \ref{fig:22assumptionComp_rangeVel}.

Beyond the fourth pass, the size of the PDF becomes large and unwieldy such that the probability densities of multiple particles dwindle to zero. Due to the logarthmic component of the entropy metric, the value of the entropy becomes ill-conditioned, resulting in the three visible gaps for the EnKF curve of Figure \ref{fig:22assumptionComp_rangeVel}. In spite of this ill conditioning, the target truth is consitent with the estimate.

Just as in Figure \ref{fig:21assumptionComp_rangeOnly}, the UKF fails to keep custody of the target during the first pass due to the unwieldiness and size of the initial state estimate. An EnKF performs even worse, by comparison. Not only is an EnKF unable to reduce the filter entropy sufficiently during the first pass, but the addition of observations in subsequent passes also does not help in refining its estimate. If it does become possible to obtain range measurements from ground-based radars or lidar sensors, the choice of filter becomes more trivial. On the other hand, the utility of minimal range information for purposes of IOD will require the use of a PGM Filter. 

\subsection{Limitations of the Proposed Framework}\label{subsec:limitations}

To summarize, our combined IOD-OD framework consists of a polynomial fitting method through several postulated positions and velocities in the topocentric reference frame in order to obtain a probabilistic initial state estimate, and the Particle Gaussian Mixture Filter in order to reduce the state uncertainty with subsequent angles-only observations. In this section, we have demonstrated the robustness of our PGM Filter, especially with relation to other, more common filters, and we have demonstrated a few different ways to obtain and filter a probabilistic initial state estimate. In this subsection, we shall discuss some of the limitations to our proposed framework, including possible improvements.

So far, our IOD framework has assumed, at a minimum, that the range bounds of the cislunar domain are roughly between 84000 km. and 550000 km., the region in which it is safe to assume that no two-body effects are present. However, not all cislunar orbits and trajectories lie within this region. About ten years ago, the Interstellar Boundary Explorer (IBEX) entered a set of highly elliptic orbits about Earth in 3:1 resonance in the Earth-Moon system.\cite{carrico2011} Researchers identified the planar mirror orbit as one of these orbits.\cite{dichmann2013} Although subjected to three-body dynamics, several regions of this orbit's trajectory lie below the defined range bounds of cislunar space. Unless we are certain that the target is within these cislunar bounds, at least during the period in which we kinematically fit through measurements, our range-bounding IOD method will require some modification.

\begin{figure}[h!]
	\centering\includegraphics[width=6in]{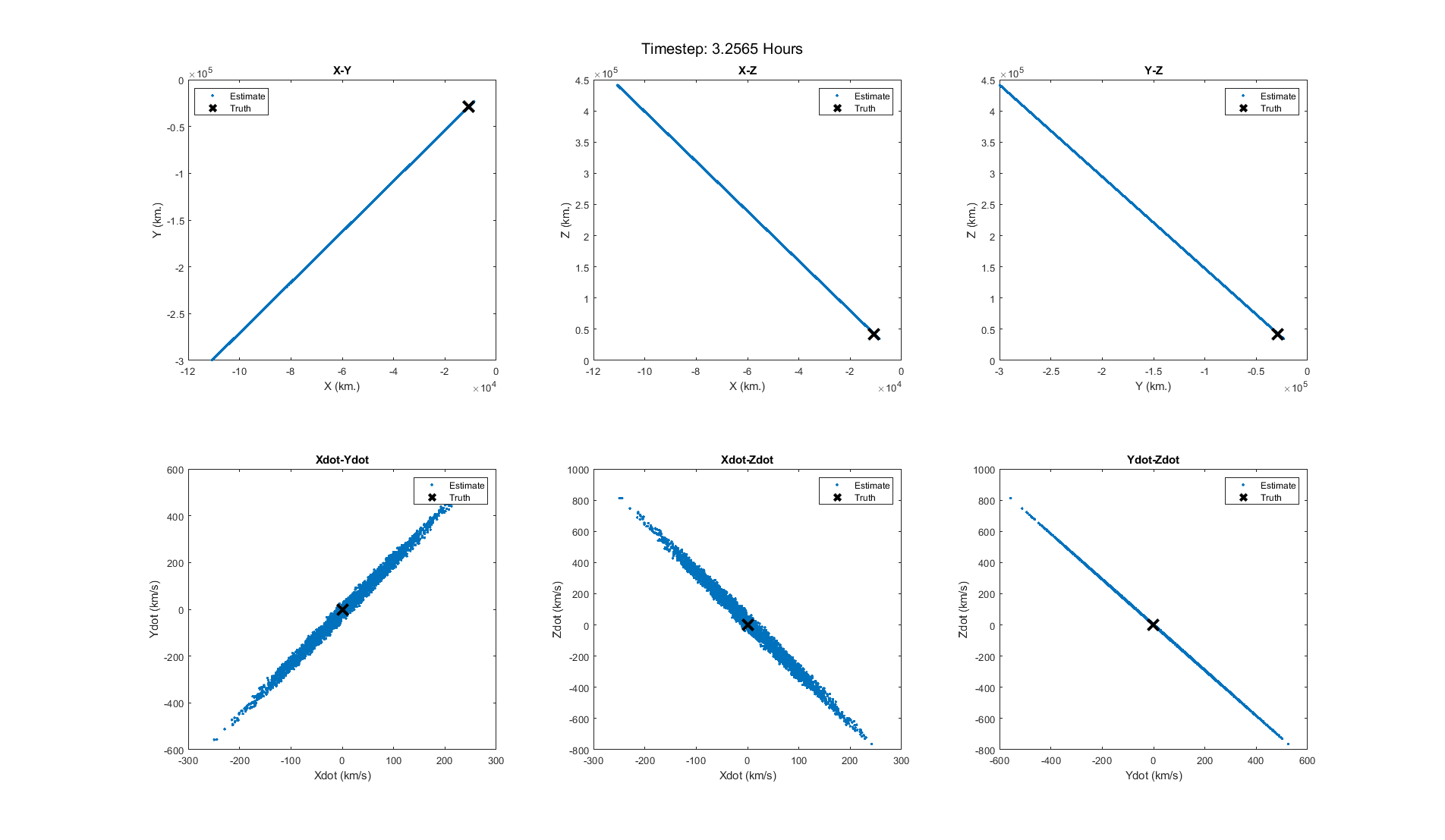}
	\caption{Initial state estimate for a target that's part of the planar mirror orbit.\cite{dichmann2013}}
	\label{fig:20limitations_IOD}
\end{figure}

Although we may be able to justify relaxing our lower range bound, allowing us to generate a consistent initial state estimate for a target in the planar mirror orbit, an issue in OD arises. As mentioned in Section \ref{subsec:OD}, the success of the PGM Filter is highly dependent on a good clustering algorithm. During the initial iterations of the PGM Filter (i.e. the first 1-2 passes), a clustering strategy that results in multiple clusters of very different sizes consistent with the observation in the measurement space may cause the PGM Filter to fail. We demonstrate this idea using the planar mirror orbit as an example. By lowering our range bound to the GEO limit (i.e. roughly 42000 km.), we obtain an initial state estimate in Figure \ref{fig:20limitations_IOD} that is consistent with the target truth, though barely.  

\begin{figure}[thpb] 
      \centering
      \begin{subfigure}{\textwidth}
            \centering
            \includegraphics[width=\linewidth]{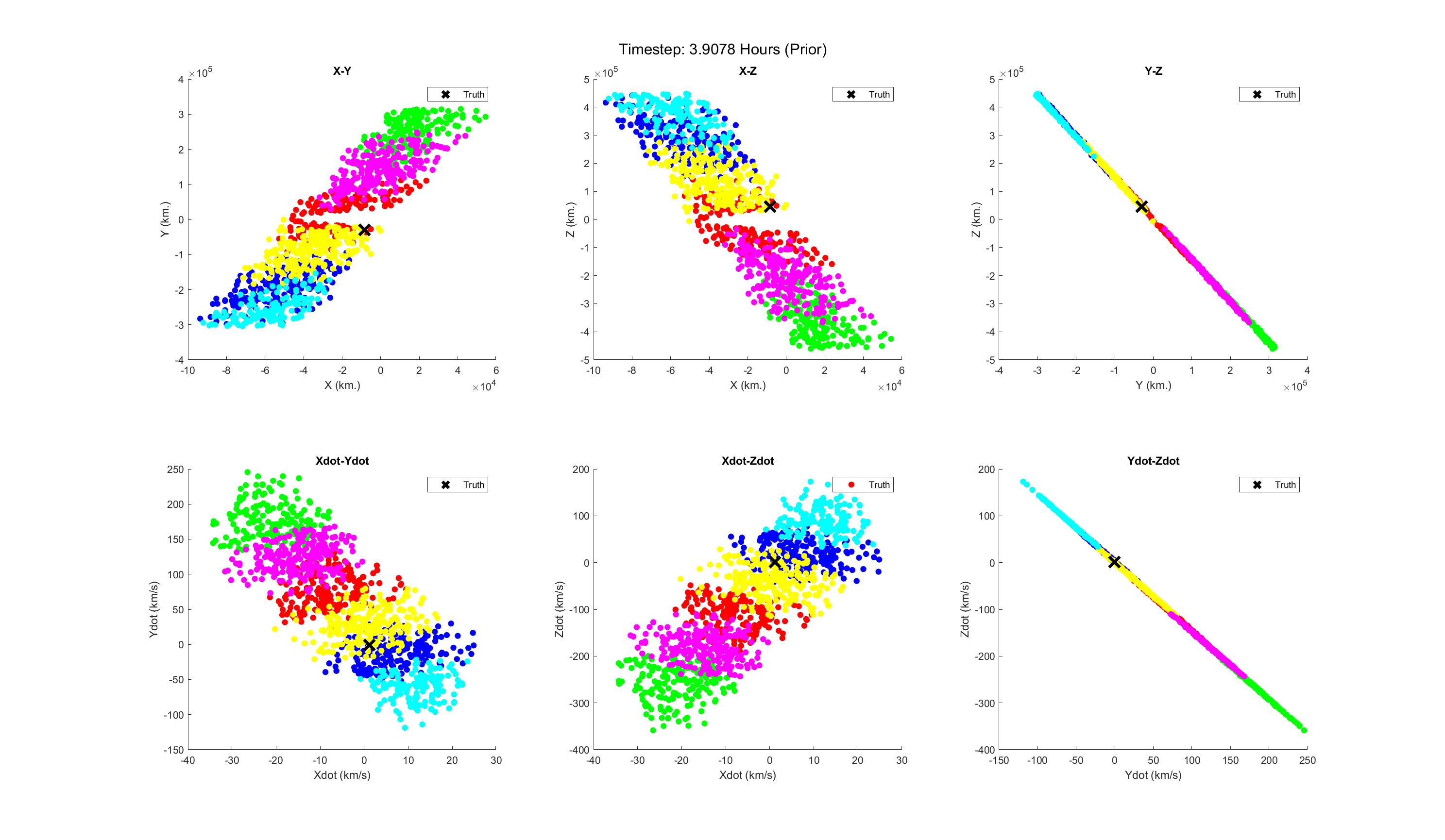}
            \caption{\textit{A Priori} estimate in the state space, once particles below the cislunar lower bound have been filtered out.} \label{fig:21a_stateSpacePrior}
      \end{subfigure}
      \vspace{0.5cm} % optional vertical spacing
      \begin{subfigure}{0.5\textwidth}
            \centering
            \includegraphics[width=\linewidth]{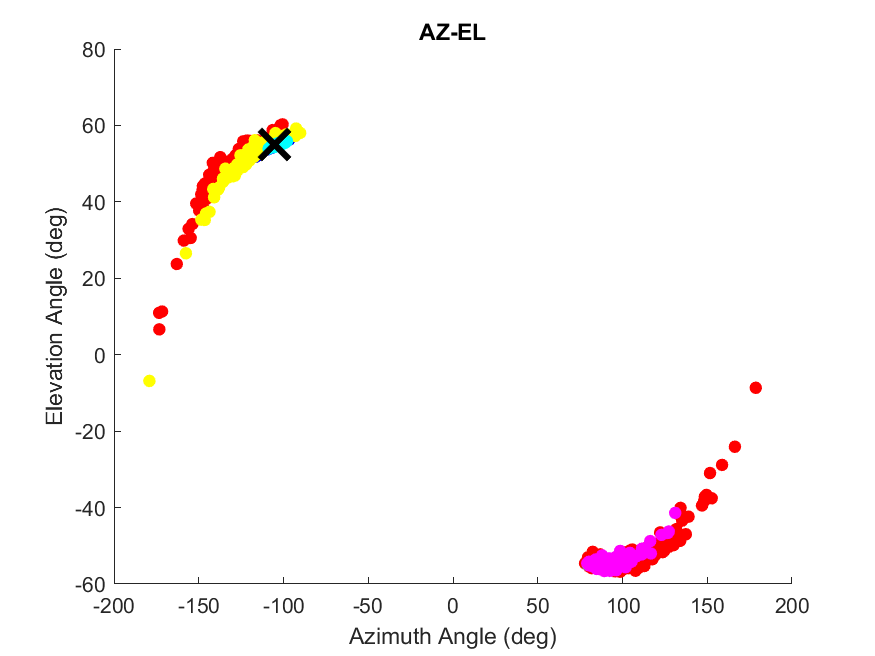}
            \caption{\textit{A Priori} estimate from Figure \ref{fig:21a_stateSpacePrior} in the measurement space.}\label{fig:21b_measSpacePrior}
      \end{subfigure}
      
      \caption{\textit{A priori} estimates for a target in the planar mirror orbit, the ensembles of which have been propagated from the initial state estimate in Figure \ref{fig:20limitations_IOD}. Multiple clusters are consistent with the observation, with greatly different measurement likelihoods.}
    \label{fig:21priors_limitation}
\end{figure}

After propagating and clustering the ensemble in Figure \ref{fig:20limitations_IOD}, we notice that due to the line-like shape of the \textit{a priori} estimate in Figure \ref{fig:21priors_limitation}, multiple clusters/GMM components, when mapped into the measurement space, are consistent with the observation, centered about the truth (and therefore, any potential observation given our sensor noise characteristics). 

\begin{figure}[h!]
	\centering\includegraphics[width=6in]{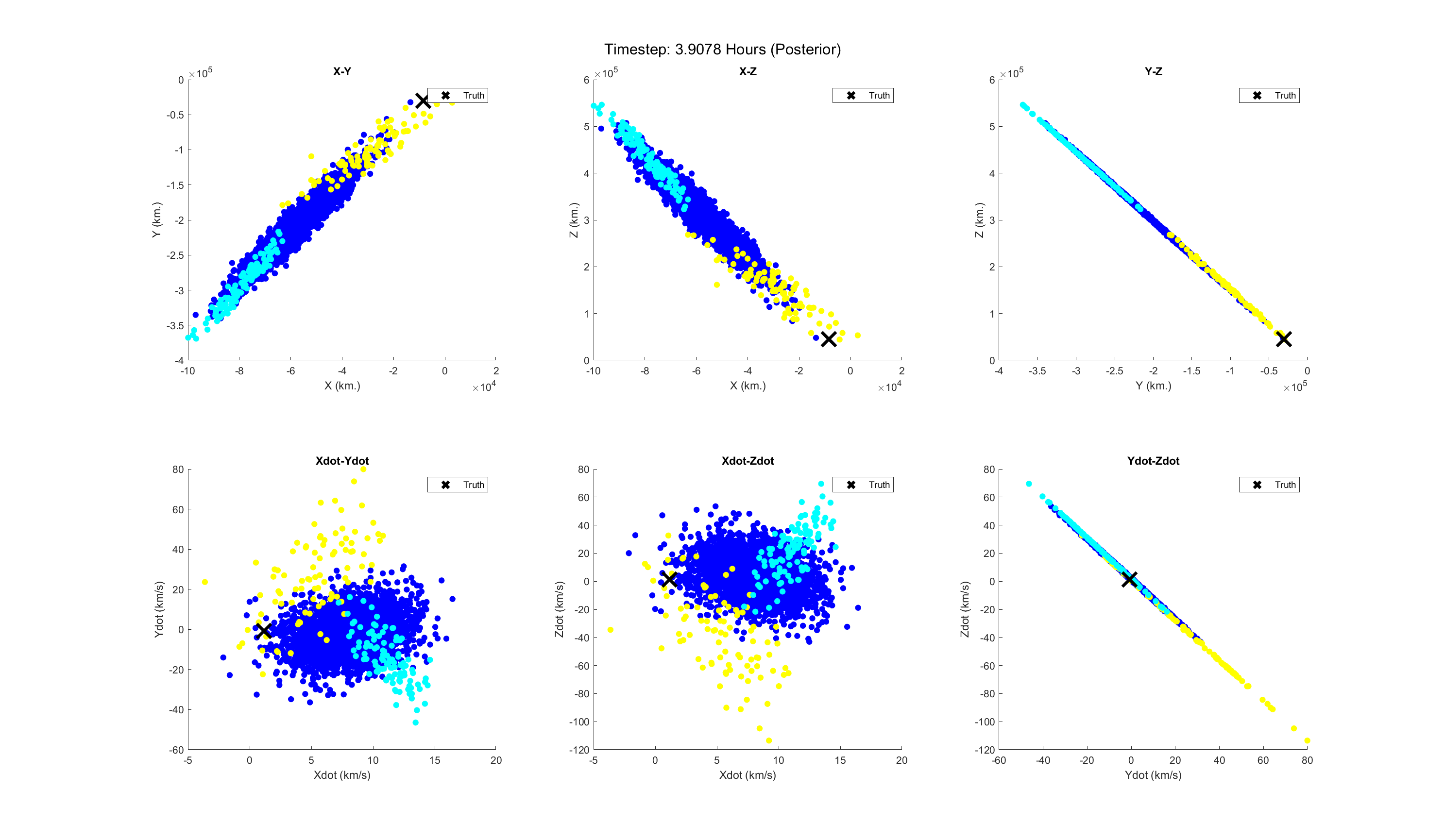}
	\caption{Posterior state estimate for a target in the planar mirror orbit after a single observation step}
	\label{fig:22posterior_limitations}
\end{figure}

Although, in the state space, one would expect the red or yellow GMM components in Figure \ref{fig:21priors_limitation} to receive the highest weight updates, the resulting measurement likelihoods are the highest with the dark blue and cyan GMM components. When translated to the measurement space, particles closer to the origin (i.e. the ground-based observer) appear more dispersed in the measurement space compared to particles further out in cislunar space due to the relatively low values of the denominators in Eq. \ref{eq8:AZ-EL}, resulting in a level of ill-conditioning. As a result, the cyan and blue clusters get most of the weight update, and our posterior state estimate in Figure \ref{fig:22posterior_limitations} starts becoming inconsistent with the truth. Since the target truth is still closer to the ground-based observer relative to the rest of the PDF, our estimate is expected to become inconsistent within just a few observation updates of the PGM Filter.

While the assumption of range bounds for orbits close to the Earth will cause the PGM Filter to fail, it is important to note that utilizing range measurements for IOD, as was described in Sections \ref{subsec:ex3} and \ref{subsec:filterComp}, will yield consistent, long-term state estimates.\cite{paranjape2025} It is also possible to choose not to fuse an observation with a target estimate and to wait until clusters of the \textit{a priori} estimate in the measurement space are easily distinguishable and non-overlapping. Alternatively, more refined clustering strategies may be developed for handling ill-conditioning or the particle spread shown in Figure \ref{fig:21b_measSpacePrior}. 

\section{Conclusions and Future Work}\label{sec:5conclusion}

In this work, we proposed and demonstrated a new IOD framework which involves fitting a polynomial or curve through series of range, azimuth, and elevation angle observations (extrapolated from noise and range inference statistics) to obtain a probabilistic initial state estimate. To address the size of our state estimate, as well as the nonlinearity of cislunar dynamics and our measurement model, we utilized the PGM Filter for OD. We demonstrated the use of our IOD-OD framework on three different orbits or trajectories in cislunar space. We focused on orbits/trajectories with moderate to severe warping due to long sensor shutoff periods, and demonstrated that the PGM framework is still capable of maintaining custody of our targets. Furthermore, the PGM Filter may render the need to perform IOD again after such long shutoffs unnecessary. 

We then demonstrated the performance of the PGM Filter relative to more commonly used filters such as UKFs and EnKFs using an entropy metric. We also provided an example trajectory and observation tasking schedule in which the UKF and EnKF will fail to maintain custody of a cislunar target while the PGM Filter will succeed. Next, we discussed three different sets of assumptions that we have and can use to generate probabilistic initial state estimates. By using the entropy metric, we demonstrated that although differing amounts of initial information or assumptions result in different precisions of state estimates in the short term, they do not vary significantly in the long term. Finally, we discussed some of the limitations of our IOD-OD framework, using the planar mirror orbit as an example. 

Although our combined IOD-OD method provides an effective target tracking framework for most objects in cislunar space, there are many possible improvements and extensions that we wish to make to this work. One of the biggest improvements we wish to make to our IOD-OD framework is the development of a clustering strategy that is robust to ill-conditioning for target estimates close to the Earth's surface. Furthermore, we wish to test this IOD-OD framework on more complex dynamics models such as the elliptical-restricted three-body problem (ER3BP), the general restricted three-body problem (R3BP), the bicircular-restricted four-body problem (BCR4BP), and even the two-body problem. Finally, we wish to study the application of this IOD-OD framework to real angles-only data, corresponding to trajectories which can significantly deviate from the underlying dynamics. 

We derive inspiration for this work from Gauss's method of orbit determination. Gauss's method, which utilizes a minimal number of assumptions about the trajectory of a target subject to Keplerian motion, extrapolates only three, consecutive short-arc measurements to obtain an initial state estimate for the second time step. Although Gauss's method is not viable in cislunar space due to the existence of non-planar motion and non-elliptic behavior, his state estimation method inspires us to develop a minimal assumption IOD method based on where on Earth we are located and where in the sky our observer points. While Gauss's method of OD is restricted to Keplerian motion and short-arc measurements, our IOD-OD framework is independent of these restrictions. Therefore, a stimulating challenge arises in developing target tracking methods with increasingly fewer assumptions in all regimes of space.

\section{Acknowledgment}
The authors of this paper are grateful to the United States Space Force (USSF) Chief Technology and Innovation Office for funding this project.
The views expressed are those of the authors and do not necessarily reflect the official policy or position of the Department of the Air Force, the Department of Defense, or the U.S. government.

\bibliographystyle{AAS_publication}   % Number the references.
\bibliography{references}   % Use references.bib to resolve the labels.

\end{document}